\documentclass[sigconf]{acmart}

\usepackage{soul}
\usepackage[raggedrightboxes]{ragged2e}
\usepackage{multirow}
\usepackage{tabularx}
\usepackage{makecell}

\AtBeginDocument{%
  \providecommand\BibTeX{{%
    \normalfont B\kern-0.5em{\scshape i\kern-0.25em b}\kern-0.8em\TeX}}}

\copyrightyear{2024}
\acmYear{2024}
\setcopyright{rightsretained}
\acmConference[ASSETS '24]{The 26th International ACM SIGACCESS Conference on Computers and Accessibility}{October 27--30, 2024}{St. John's, NL, Canada} \acmBooktitle{The 26th International ACM SIGACCESS Conference on Computers and Accessibility (ASSETS '24), October 27--30, 2024, St. John's, NL, Canada} \acmDOI{10.1145/3663548.3675617} \acmISBN{979-8-4007-0677-6/24/10}

\begin{document}

\title{Audio Description Customization}

\author{Rosiana Natalie}
\affiliation{
    \institution{Singapore Management University}
    \country{Singapore}
    }
\email{rnatalie.2019@phdcs.smu.edu.sg}

\author{Ruei-Che Chang}
\affiliation{
    \institution{University of Michigan}
    \city{Ann Arbor}
    \state{MI}
    \country{USA}
    }
\email{rueiche@umich.edu}

\author{Smitha Sheshadri}
\affiliation{
    \institution{Singapore Management University}
    \country{Singapore}
    }
\email{smithas.2022@phdcs.smu.edu.sg}

\author{Anhong Guo}
\affiliation{
    \institution{University of Michigan}
    \city{Ann Arbor}
    \state{MI}
    \country{USA}
    }
\email{anhong@umich.edu}

\author{Kotaro Hara}
\affiliation{
    \institution{Singapore Management University}
    \country{Singapore}
    }
\email{kotarohara@smu.edu.sg}

\begin{abstract}
  Blind and low-vision (BLV) people use audio descriptions (ADs) to access videos. However, current ADs are unalterable by end users, thus are incapable of supporting BLV individuals' potentially diverse needs and preferences. This research investigates if customizing AD could improve how BLV individuals consume videos. We conducted an interview study (Study 1) with fifteen BLV participants, which revealed desires for customizing properties like \textit{length, emphasis, speed, voice, format, tone}, and \textit{language}. At the same time, concerns like interruptions and increased interaction load due to customization emerged. To examine AD customization's effectiveness and tradeoffs, we designed CustomAD, a prototype that enables BLV users to customize AD content and presentation. An evaluation study (Study 2) with twelve BLV participants showed using CustomAD significantly enhanced BLV people’s video understanding, immersion, and information navigation efficiency. Our work illustrates the importance of AD customization and offers a design that enhances video accessibility for BLV individuals.
\end{abstract}


\begin{teaserfigure}
  \includegraphics[width=\textwidth]{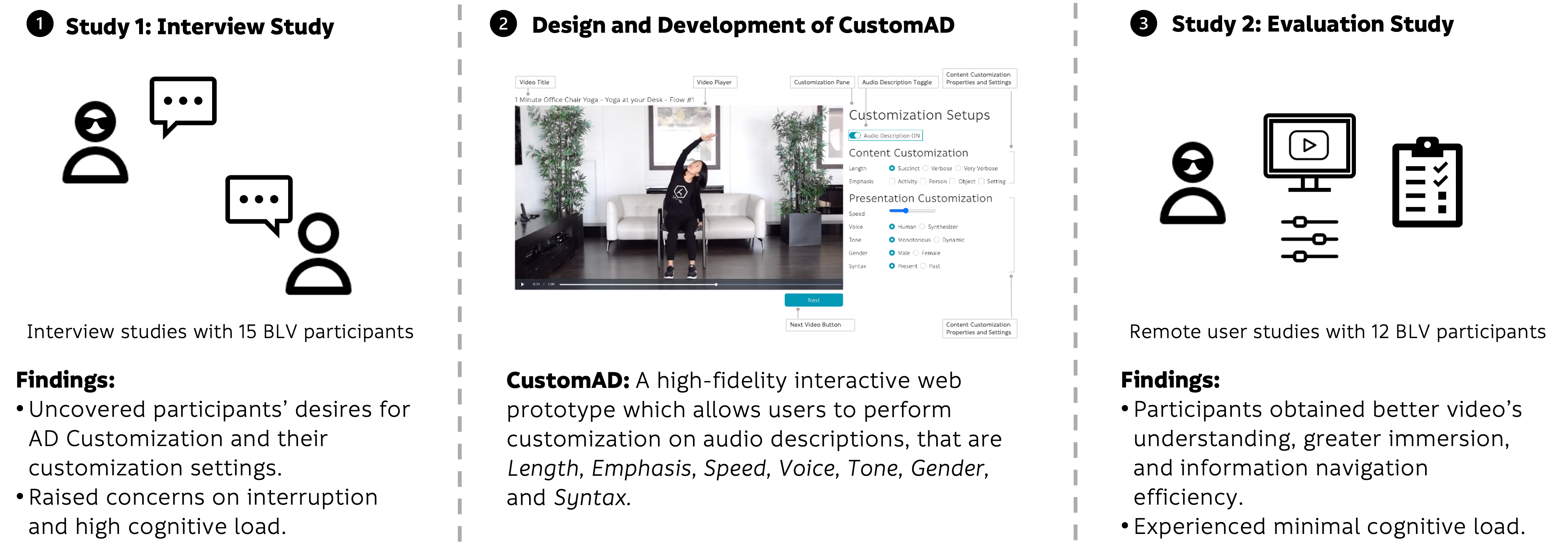}
  \caption{Our research on the customization of audio descriptions consists of three parts: 1) An interview study with 15 BLV participants to uncover their desires and preferences in audio descriptions customizations, 2) the design and development of CustomAD, a high-fidelity prototype that reflects the audio description customization preferences that emerged from the interview study, and 3) an evaluation study on the effectiveness and trade-offs of audio description customization.}
  \Description[Teaser Figures]{Summary of our research. The figure consists of three columns. The first column where we explain the interview study. The second column is where we explain our design and development of CustomAD, with screenshots of the interface. The third column summarises the evaluation study.}
  \label{fig:teaser}
\end{teaserfigure}

\maketitle

\section{Introduction}
Blind and low-vision individuals (BLV) depend on audio descriptions (ADs), verbal narrations of visual content, to comprehend videos ~\cite{fryer2016introduction,w3c-wai, netflixADGuide,walczak2018vocal,w3c}. 
Guidelines for authoring ADs outline the qualities that good descriptions should satisfy like descriptiveness and succinctness \cite{acb_guideline, DCMP, netflixADGuide, youdescribe2020}. 
A growing body of research focuses on enhancing the efficiency of AD production while meeting these quality standards, either by supporting manual authoring with technologies \cite{DCMP,kobayashi2009providing,thompson2019,ableplayer,natalie2020viscene,natalie2021efficacy, gagnon2010computer, gurari2020captioning, natalie2023supporting, yuksel2020human,yuksel2020increasing,pavel2020rescribe} or by automating the process  \cite{campos2020cinead, wang2021toward, lei2020mart, ni2022expanding}. 
However, the premise of these efforts is that offering one good description for each video will satisfy everyone, which may overlook diverse preferences and needs of BLV individuals when consuming ADs. 
While such generic ``one-size-fits-all'' ADs are generally accepted by BLV individuals ~\cite{chmiel2022homogenous}, they may not be optimal.

Despite the increasing interest in technological solutions for AD production, studies aimed to understand the diversity in AD needs and the demand for personalized ADs remain relatively sparse. 
Prior work has mentioned the inadequacy of generic ADs in fully meeting the needs of BLV individuals and has called for more investigations \cite{jiang2023beyond, wang2021toward}; still, the investigations were left as future work.
Most relevant to the current work are recent studies by Jiang \textit{et al.} and Chmiel and Mazur, which offer evidence showing that AD needs among BLV individuals indeed vary depending on the contexts in which videos are consumed ~\cite{jiang2024it} and the levels of visual impairments ~\cite{chmiel2022homogenous}. 
While informative, several critical questions remain unanswered. 
For instance, what specific characteristics of ADs are essential for personalization, and how can we effectively cater to such diverse needs?

To extend the prior study in understanding BLV individuals' AD preferences and explore ways for AD personalization, we ask: 
\textit{Is enabling BLV end-users to customize ADs desirable as a way to cater to diverse needs, and if so, what customization properties are perceived to be meaningful and important?} 
\textit{How could we design a tool to support AD customization?} 
And, \textit{how does AD customization affect video consumption, and how do BLV individuals perceive its impact?}

To address the first question, we conducted remote semi-structured interviews (Study 1) with fifteen BLV individuals.
Participants first watched seven different types of videos with ADs, such as instructional videos and documentaries. 
We then asked questions to determine their interest in AD personalization, whether customization is a favorable approach to personalization, and which AD properties they consider important to customize. 
Participants' responses suggested that AD customization could improve experience in consuming videos, enhance AD clarity, and increase video immersion and efficiency of information consumption. 
Our findings also revealed that both content-related properties (\textit{e.g.,} \textit{length}, \textit{emphasis}) and presentation-related characteristics (\textit{e.g.,} \textit{speed}, \textit{tone}) could improve how BLV individuals consume ADs.

Motivated by the results of Study 1, we designed and developed CustomAD, a high-fidelity web-based prototype that enables users to customize both the content and presentation properties of ADs. 
Users could adjust content properties such as \textit{length} and \textit{emphasis}, as well as presentation properties including \textit{speed}, \textit{voice}, \textit{tone}, \textit{gender}, and \textit{syntax} by controlling values in form elements (Fig.~\ref{fig:prototype_figure}). 
The CustomAD interface was designed for accessibility, allowing BLV users to navigate using keyboard shortcuts.
Any changes to customization settings were reflected immediately in the ADs, allowing the user to assess if the customized AD meets their liking.

Using CustomAD as an apparatus, we conducted a remote evaluation study (Study 2) with twelve BLV participants to investigate the effectiveness of AD customization and the tool's usability. 
The study was a two (\textit{with} \textit{vs.} \textit{without} customization) by three (\textit{entertainment}, \textit{explainer}, and \textit{tutorial} videos) within-subjects design. 
Participants were asked to use CustomAD to watch six videos, two in each video type. 
Each video was accompanied by diverse set of AD versions for customization authored by a professional. 
To measure the effect of customization on video comprehension, we asked participants to identify specific information from the ADs. 
We assessed their accuracy and the time taken to complete these tasks. 
Additionally, we collected subjective metrics to evaluate the tool's usability (\textit{e.g.,} Likert-scale responses measuring perceived usefulness and NASA TLX questionnaire to measure customization task load). 
The study concluded with a brief interview. 
Our findings showed that that participants' accuracy in performing the information identification task was significantly higher with customization. 
However, they took longer to complete tasks, as they spent more time interacting with the customization interface. Subjective data suggested that using CustomAD was easy and comfortable, with a manageable cognitive load.

In summary, our work makes the following contributions:
\begin{itemize}
    \item Findings from the interview study with fifteen BLV individuals that uncover the desire for customizing ADs and important customization properties.
    \item The design and development of CustomAD that enables customization of ADs to suit BLV individuals' preferences and needs.
    \item Empirical findings from the evaluation study with twelve BLV participants demonstrating the effectiveness of AD customization.
\end{itemize}

\section{Related Work}
\subsection{Audio Descriptions}
Legislation in many countries mandate provision of audio descriptions (ADs) across various media, including television, cinema, and digital platforms~\cite{leung2018audio, cronin1990development, wcag2008, fcc2020, greening2007accessibility, BlindCitizenAustralia, vera2006translating}).
For example, in the US, CVAA Title 2~\cite{fcc2020} requires major broadcast and cable networks to make online videos accessible. The UK’s 1996 Broadcasting Act specify minimum numbers or percentages of programs and their duration that must be made accessible. As the response the broadcating act, UK's TV broadcasters voluntarily have offering up to 20\% of their airtime with ADs to further enhance accessibility~\cite{katsarova2018audiovisual}. These efforts have increased the availability of ADs, in turn improved how BLV individuals consume traditional visual media~\cite{leung2018audio}. 

AD authoring guidelines, developed by organizations like the American Council of the Blind~\cite{acb_guideline}, DCMP~\cite{DCMP}, ADLab~\cite{adlabGuideline}, and Netflix~\cite{netflixADGuide}, also, prior works on visual descriptions (\textit{e.g.,}~\cite{stangl2020person, stangl2021going, natalie2021efficacy, wang2021toward})  provide detailed recommendation on how to craft ADs that are useful for BLV individuals.
These guidelines and research suggest what to include in ADs (\textit{e.g.,} focusing on the important visual content and progressing from the general to the specific~\cite{stangl2020person, stangl2021going}, avoiding descriptions that can be inferred from the sound~\cite{acb_guideline, netflixADGuide}, and providing right amount of information~\cite{natalie2021efficacy, natalie2023supporting, wang2021toward}), placement (\textit{e.g.,} ensure they do not overlap with dialogues~\cite{acb_guideline, ndr, mac, DCMP, adp, w3c, wcag}), and style of delivery (\textit{e.g.,} match the tone~\cite{acb_guideline, ndr, adp} and vocabulary of the source video, use active voice~\cite{acb_guideline, mac, DCMP, adp}, acceptable speed~\cite{acb_guideline, DCMP, adp}, and avoid editorializing~\cite{acb_guideline}).
WCAG 2.0's Success Criterion 1.2.7~\cite{WCAG_SC12-7} advocates for the implementation of extended ADs, which temporarily pause audio and video in the original content to deliver important visual details, especially when the natural pauses in dialogue are not long enough for a detailed description.

Prior research has explored technological solutions to facilitate AD authoring while adhering to these guidelines. 
For example, Pavel \textit{et al.} introduced Rescribe, a system that helps novice authors craft inline extended ADs that seamlessly integrate with other video contents ~\cite{pavel2020rescribe}. 
Chang \textit{et al.} developed Omniscribe~\cite{chang2022omniscribe}, a tool is designed for both authoring and delivering ADs tailored to 360\textdegree{} content. 
Chang's approach enhances AD immersion by incorporating spatial audio, vibrations to signal scene transitions, and tracking head movements, introducing ways to present AD effectively for BLV individuals to enjoy 360\textdegree{} videos.

Although these guidelines and technologies advance the goal of making videos more accessible, they implicitly emphasize the existence of an ideal AD that is presumed to suit each video, and so creating a ``one-size-fits-all'' AD could meet the needs of every BLV individual. 
This assumption, however, overlooks the diverse preferences, needs, and abilities within BLV individuals. 
This paper explores the variations in AD preferences and the specific properties that BLV individuals wish to customize. 
We also evaluate how such customization impacts video comprehension and immersion.

\subsection{Audio Descriptions Preferences}
Characteristics of ADs identified in prior work as crucial for BLV individuals may also be suitable for customization
~\cite{campos2020cinead, liu2021makes, pavel2020rescribe, wang2021toward, yuksel2020human, natalie2020viscene, natalie2021efficacy, natalie2023supporting, jiang2022co}. 
For example, Jiang \textit{et al.}, in developing Accessible AD, found that BLV individuals appreciate details related to the characters, background settings, and actions in videos~\cite{jiang2022co}. 
Yuksel \textit{et al.} showed that BLV individuals seek precise direction and measurements, and emphasizing such information in cooking videos to be beneficial~\cite{yuksel2020human}. 
In developing and evaluating a tool that supports authors to create ADs, Pavel \textit{et al.} found BLV individuals value ADs that offer extensive details without overlapping with the original audio track~\cite{pavel2020rescribe}. 
The relevance of these characteristics like content emphasis and level of detail would vary depending on the audience and context. 
Thus, they provide a starting point for examining variations in AD preferences and exploring potential customization.

Some studies have explored the diverse preferences and information needs of BLV individuals for visual media~\cite{chmiel2016researching, chmiel2022homogenous, jiang2023beyond, jiang2024it, stangl2020person, stangl2021going, lopez2018audio, leung2018audio}. 
The work by Stangl revealed the varied description needs of BLV individuals, such as contexts of image use, types of sharing platforms, and the goal of the information sought~\cite{stangl2020person, stangl2021going};
though their focus was on image descriptions, it is plausible such variations in needs exist for videos's ADs, too.
In fact, a study by Lopez \textit{et al.}~\cite{lopez2018audio} suggested that BLV individuals demand personalized ADs that are tailored to their unique abilities and interests. 
Studies that are directly relevant to the current work include recent research by Jiang \textit{et al.}~\cite{jiang2024it} and Chmiel and Mazur ~\cite{chmiel2022homogenous}. 
Jiang \textit{et al.} revealed BLV individuals' preferences on levels of details and output modalities for different types of videos ~\cite{jiang2024it}.
Chmiel and Mazur found that, while BLV individuals generally preferred ADs that adhere to existing guidelines, variations in their residual vision affected preferences on characteristics like character naming~\cite{chmiel2022homogenous}.
While these studies provide evidence of differences in individual preferences for AD delivery, they primarily focused on whether viewing scenarios and levels of visual impairment affect AD preferences.
To complement and extend these findings, we conduct an interview study to explore the preferences for low-level characteristics of ADs, such as length, emphasis, speed, and voice, identifying which features are more desired for AD customization. 
Furthermore, through the design and evaluation of CustomAD, we investigate the objective effectiveness and subjective usability of end-user customization of ADs, providing insights into user interactions for personalized ADs.

\subsection{Customization of User Interfaces} 
Previous research in Human-Computer Interaction (HCI) has highlighted the benefits of enabling end-users to customize various aspects of their digital environments. 
This includes customization of menus and toolbars~\cite{bunt2007supporting, hurst2007dynamic, de2013firefixia}, interface layouts~\cite{saati2005towards,sundar2010personalization, wu2009biogps, ponsard2016anchored, damian2011individualized, sung2010social, gajos2004supple}, content~\cite{pawlowski2010basic, halsted2002eclipse}, and information visualization~\cite{abla2010customizable, beam2014personalized, kennedy2012taxonomy}.
Customization has shown to be effective in supporting the unique needs of individuals with disabilities, too.
For instance, blind users have adjusted options for screen readers~\cite{NVDA,JAWS,VoiceOver}, online discussion forums~\cite{sunkara2023enabling}, navigation tools~\cite{kuriakose2022tools}, audio books~\cite{morley1998digital}, graphical user interfaces~\cite{gajos2004supple}, and augmented reality~\cite{montagud2020culture} to better suit their needs.
The past work consistently supports that customization improves digital tools' usability and accessibility~\cite{herskovitz2023hacking, stangl2020person, stangl2021going}.
Thus, designing a tool that gives BLV individuals autonomy and control to customize the content and presentation of ADs could be a viable design direction for AD personalization.
In this research, we explore what BLV individuals want to customize in ADs beyond basic AD toggling feature that existing video platforms like YouTube~\cite{youtube}, Netflix~\cite{netflix}, Disney+~\cite{disney}, and Amazon Prime~\cite{amazonPrime} provide.
Our study is the first to design, develop, and evaluate the effect of AD customization on the video-watching experience of BLV individuals.

\section{Study 1: Interview Study Method}
We conducted remote semi-structured interviews with BLV individuals to investigate whether AD customization is viable for accommodating their diverse AD needs. The study also explored what customization properties are perceived as essential.

\begin{table*}[]
\small
\begin{tabular}{llllllll}
\hline
\textbf{ID} &
  \textbf{Gender} &
  \textbf{Age} &
  \textbf{\begin{tabular}[c]{@{}l@{}}Primary  Occupation\end{tabular}} &
  \textbf{\begin{tabular}[c]{@{}l@{}}Level of Vision\end{tabular}} &
  \textbf{\begin{tabular}[c]{@{}l@{}}Visual Onset\end{tabular}} &
  \textbf{Diagnosis} &
  \textbf{\begin{tabular}[c]{@{}l@{}}Frequency of Watching\end{tabular}} \\ \hline
\textbf{P1$\ast$} &
  Male &
  34 &
  Freelance &
  Blind &
  Acquired &
  \begin{tabular}[c]{@{}l@{}}Cataract \& Retinal \\ Detachment\end{tabular} &
  Everyday \\
\textbf{P2} &
  Male &
  25 &
  \begin{tabular}[c]{@{}l@{}}Digital accessibility \\ specialist\end{tabular} &
  Low Vision &
  Congenital &
  Glaucoma &
  Everyday \\
\textbf{P3} &
  Female &
  59 &
  Tour guide &
  Blind &
  Congenital &
  Cataract \& Glaucoma &
  Once a week \\
\textbf{P4$\ast$} &
  Male &
  26 &
  Trainer and consultant &
  Low Vision &
  Congenital &
  Retinal Dystrophy &
  Undetermined \\
\textbf{P5$\ast$} &
  Male &
  42 &
  Executive &
  Low Vision &
  Acquired &
  Retinitis Pigmentosa &
  Once a week \\
\textbf{P6$\ast$} &
  Male &
  41 &
  Technology analyst &
  Low Vision &
  Congenital &
  Retinitis Pigmentosa &
  Everyday \\
\textbf{P7} &
  Female &
  34 &
  Receptionist &
  Low Vision &
  Congenital &
  Phthisis Bulbi &
  Everyday \\
\textbf{P8$\ast$} &
  Male &
  62 &
  Retired &
  Low Vision &
  Acquired &
  \begin{tabular}[c]{@{}l@{}}Congenital Sclerocornea \& \\ Glaucoma\end{tabular} &
  Everyday \\
\textbf{P9} &
  Male &
  66 &
  Retired &
  Blind &
  Acquired &
  \begin{tabular}[c]{@{}l@{}}Congenital Cataracts \& \\ Glaucoma\end{tabular} &
  Everyday \\
\textbf{P10$\ast$} &
  Female &
  50 &
  Senior manager &
  Blind &
  Acquired &
  \begin{tabular}[c]{@{}l@{}}Central Retinal \\ Artery Occlusion\end{tabular} &
  Once a week \\
\textbf{P11$\ast$} &
  Female &
  54 &
  Part time tour guide &
  Blind &
  Congenital &
  Retinal Detachment &
  Undetermined \\
\textbf{P12$\ast$} &
  Female &
  39 &
  Restaurant server &
  Low Vision &
  Congenital &
  Maculopathy &
  Everyday \\
\textbf{P13} &
  Male &
  40 &
  Call center agent &
  Low Vision &
  Congenital &
  Cone Exstrophy &
  Everyday \\
\textbf{P14$\ast$} &
  Female &
  26 &
  Administrative assistant &
  Blind &
  Congenital &
  Retinopathy of Prematurity &
  Everyday \\
\textbf{P15$\ast$} &
  Female &
  29 &
  Civil servant &
  Blind &
  Acquired &
  Glaucoma &
  Once a week \\
\textbf{P16$\ast$} &
  Male &
  24 &
  Student &
  Low Vision &
  Congenital &
  Retinitis Pigmentosa &
  Everyday \\
\textbf{P17$\ast$} &
  Female &
  27 &
  Coach and facilitator &
  Low Vision &
  Congenital &
  Aniridia and Glaucoma &
  Everyday \\ \hline
\end{tabular}
\vspace{0.5pc}
\caption{Demographic information of the participants for Study 1 and Study 2. P1 to P15 participated in Study 1. Participants annotated with an asterisk ($\ast$) participated in Study 2 (ten participated in both studies). For level of vision, Blind indicates participants with total blindness and Low Vision represents participants with low-vision and legally blind participants.}
\vspace{-1.5pc}
\label{tab:participant_demographic}
\end{table*}

\subsection{Participants}
We recruited N=15 participants through snowball sampling \cite{goodman1961snowball}. All the participants completed the demographic questionnaire before the study. Among the fifteen participants (7 female, 8 male), six had congenital visual impairments, and nine acquired their visual impairments later in their life. Seven participants were totally blind, and eight were low-vision. The average age of the participants was 42 years old (SD = 13.63, Md = 40). See the demographic information in Table~\ref{tab:participant_demographic}.

\begin{figure*}
    \centering
    \includegraphics[width=14.5cm]{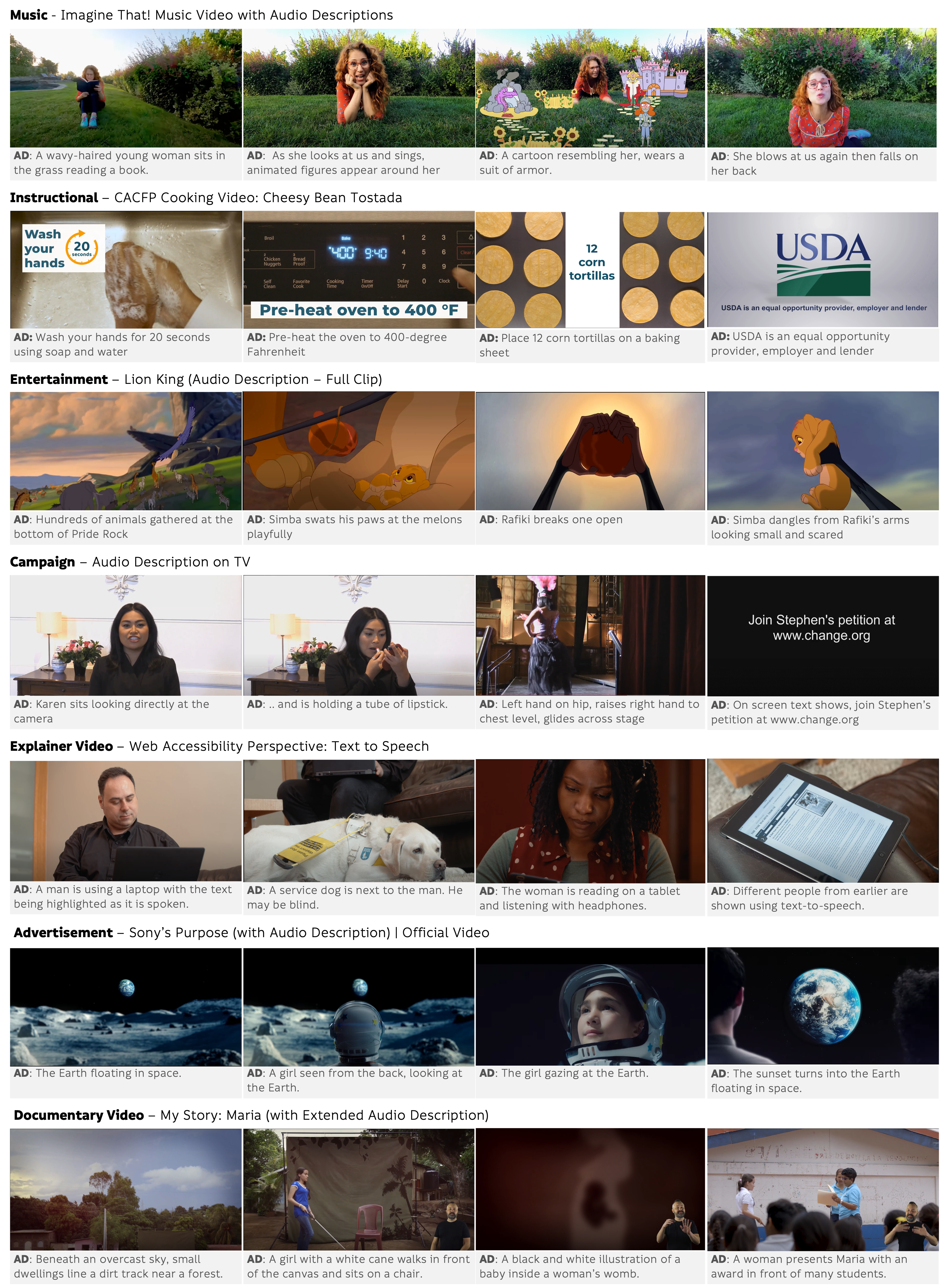}
        \caption{Videos and their ADs used in Study 1. Each row represent a video. Video types are: \textit{music}, \textit{instructional}, \textit{entertainment}, \textit{campaign}, \textit{explainer}, \textit{advertisment}, and \textit{documentary}. Each participant watched all seven videos to increase their awareness of different AD contents and styles.}
    \Description[Screenshots of the video apparatus.]{Screenshots of several scenes in the videos we used in our Study 1. The figure is presented in a grid format. Each row represents one video type. Types of the videos, from top to bottom, are music, instructional, entertainment video, campaign, explainer, advertisement, and documentary. We show four example scenes for each video. Below each screenshot, we show its corresponding audio description.
    }
    \label{fig:formative_video}
\end{figure*}

\subsection{Procedure}
We conducted the study remotely over Zoom or FaceTime. 
Each session consisted of a video-watching activity and an interview session in which we discussed customization viability and preferences. 
To expose the participants to various videos with ADs and to establish a common ground for the subsequent interview about customization preferences, we asked the participants to watch seven videos of different types (Fig.~\ref{fig:formative_video}). 
In the interview, we asked participants about their attitudes toward AD customization and their thoughts on how different customization properties could assist them in consuming ADs. 
For each property, we asked participants to rate their agreement to the statement, \textit{``The customization can help me to consume AD more effectively,''} on a 5-point Likert scale ($1 =$ Strongly Disagree to $5 =$ Strongly Agree).
We also asked the participants for their rationale for the scores. 
Additionally, we invited participants to suggest any customization properties not covered by our customization properties list described in Section ~\ref{sec:customization_properties} that they believed could also enhance AD consumption.

Our study procedure was approved by our institution’s IRB. 
Before the study started, the participants gave consent to participate verbally. 
Each session lasted for about two hours.
We compensated participants with \$40 for their participation upon completing the session.

\begin{table*}[]
\small
\begin{tabular}{lp{5.5cm}p{7cm}}
\hline
\textbf{Properties} &
  \textbf{Customization Properties Explanations} &
  \textbf{Settings \& Examples} \\ \hline
\multicolumn{3}{l}{\textbf{Content Customization}} \\ \hline
Length   \cite{wang2021toward,natalie2021efficacy,natalie2023supporting} &
  Length of the AD, which also acts as the proxy of the amount of information in the AD. The more verbose the AD, the more information is covered. & \textbf{Settings:} Succinct\textasciicircum{}, Verbose, Very Verbose \newline 
  Succinct: \textit{Elsa looks at Anna sadly.}\newline Verbose:  \textit{In a ballroom crowded with richly-dressed men and women, Elsa, a young white woman with blond hair, walks away from Anna, a young, red-haired white woman. Both wear gowns, Elsa’s more conservative. She also wears a tiara. Beside Anna is Hans, a formally-dressed brunet white man. Anna rushes towards Elsa, reaching for her hand. She pulls off one of Anna’s gloves.} \\ \hline
Emphasis\cite{stangl2020person,stangl2021going} &
  Information category users could focus on in AD. We allow this property to be empty (\textless{}null\textgreater{}), so the AD will have a balanced information emphasis. & \textbf{Settings:} \textless{}null\textgreater{}\textasciicircum{}, Activity, Person, Object, Setting \newline 
  Activity $\times$ Succinct: \textit{Elsa sadly looks at Anna.}\newline Object $\times$ Verbose: \textit{In the ballroom, Elsa walks away from Anna. Elsa wears a conservative gown with a tiara, while Anna wears a tiara as well. Anna rushes towards Elsa, swiftly removing her glove.}\\ \hline
\multicolumn{3}{l}{\textbf{Presentation Customization}} \\ \hline
Speed \cite{adp, acb_guideline, DCMP, bragg2021expanding} &
  AD speed in the range of 0.25 to 2, with the increment of 0.25. Values greater than 1.0x denote users speeding up the video. Values below 1.0x indicate users slowing down the video. The default value is 1.0x & \textbf{Settings:} A slider is provided for selecting speeds from 0.25x to 2.0x, with increments of 0.25. Default is 1.0x$\ast$ 
   \\ \hline
Voice \cite{ndr,adp,acb_guideline,3PlayMedia_BeginnerGuideline}&
  Voice that reads out AD, which could be human voice or synthesizer voice. & \textbf{Settings:} Human\textasciicircum{}, Synthesizer
   \\ \hline
Tone \cite{ndr,adp,acb_guideline} &
  Tone of AD, which could be monotonous or dynamic. & \textbf{Settings:} Monotonous\textasciicircum{}, Dynamic 
   \\ \hline
Gender \cite{netflixADGuide} &
  Gender of the voice that reads out AD.  & \textbf{Settings:} Male\textasciicircum{}, Female 
   \\ \hline
Syntax \cite{adp,acb_guideline, DCMP, mac} &
  Grammatical syntax of the AD. AD could be narrated in present or past tense. & \textbf{Settings:} Present\textasciicircum{}, Past \newline
  Present: \textit{They look at each other sadly. Hiding her ungloved hands, Elsa walks to the door, her cape trailing behind her. Other guests stare. Guest turn to look.} \newline Past: \textit{They looked at each other sadly. Hiding her ungloved hands, Elsa walked to the door, her cape trailing behind her. Other guests stared. Guest turned to look.} \\ \hline

Format$\ast$ \cite{ndr, adp, acb_guideline, DCMP, mac, wcag} &
  Inline or extended. Inline ADs are designed to fit naturally within the existing gap between dialogues in a video. Extended ADs are longer and require pausing the video to fit their full duration. &
  - \\ \hline

  Language$\ast \ast$ &
  Customize the language of the AD to a language that the BLV viewer understands. &
  - \\ \hline
\end{tabular}
\caption{Summary of customization properties explored in and emerged from Study 1. Settings \& Examples column describes the options supported by CustomAD (options marked with `\textasciicircum{}' are default). ($\ast$) In CustomAD, \textit{format} was automatically adjusted depending on the \textit{length} properties (\textit{i.e.,} when ADs were too long to fit in the available pause, which was usually the case for the \textit{Verbose} and \textit{Very Verbose} ADs, they were presented in the \textit{extended} format). ($\ast \ast$) \textit{Language} property emerged from the interview. We did not implement it in CustomAD because the videos were in English and all participants were fluent in English.}
\vspace{-1.5pc}
\label{tab:customAD_customization_properties}
\end{table*}

\subsection{Apparatus}
\subsubsection{Videos}
To provide participants with a comprehensive and diverse experience with AD, we selected videos with a wide range of types for the video-watching activity. 
These video types included: 
(a) music video of a song, ``Imagine That''~\cite{imagineThat}, 
(b) instructional videos on making a cheese tostada~\cite{cheeseTostada}, 
(c) an entertainment video, a clip from the movie ``Lion King''~\cite{lionKing}, 
(d) a campaign video promoting ADs on TV in Australia~\cite{campaignAustralia}, 
(e) an explainer video on text-to-speech~\cite{campaignAustralia}, 
(f) an advertisement video showcasing Sony products~\cite{sonyAds}, and 
(g) a documentary video featuring Maria, a visually impaired girl living in Nicaragua~\cite{mariaDocument}. 
These videos were selected from reputable sources and featured high-quality AD crafted by experts. 
See Fig.~\ref{fig:formative_video} for the screenshots of each video's scenes and their corresponding AD.

\subsubsection{Customization Properties}\label{sec:customization_properties}
To understand customization properties preferred by BLV individuals, we compiled a list of properties by reviewing existing AD guidelines \cite{acb_guideline,adp,DCMP,netflixADGuide, 3PlayMedia_BeginnerGuideline, mac, ndr}. 
We identified six primary customization properties: \textit{speed, voice, format, tone, gender,} and \textit{syntax} which primarily pertain to the presentation aspects of ADs. 
Drawing from prior research, (\textit{e.g.}, \cite{wang2021toward, natalie2021efficacy, stangl2020person, stangl2021going}), we also explored opportunities to enhance and customize the ADs content themselves. 
We examined the potential benefits of customizing \textit{length} and \textit{emphasis} as proxies to adjust the verbosity of information and focus on desired content, respectively. 
We summarized the list of customization properties in Table~\ref{tab:customAD_customization_properties}.

\section{Study 1: Interview Study Result}

We used a content analysis to iteratively code and analyze interview transcripts~\cite{drisko2016content, hruschka2004reliability}. 
We transcribed the interview recordings using Whisper \cite{radford2023robust}. 
The first author reviewed all transcripts and generated the initial codebook through open and axial coding. 
Two authors then independently coded two transcripts, achieving a Cohen’s Kappa score ($\kappa$) of 0.81. 
The two authors resolved the disagreement and coded three additional interview transcripts ($\kappa = 0.86$). 
Three more transcripts were coded, resulting in a $\kappa = 0.94$ agreement. 
The two authors resolved disagreements and finalized the codebook. 
One researcher used the final codebook to code the remaining interview transcripts.

\begin{figure*}
    \centering
    \includegraphics[width=15cm]{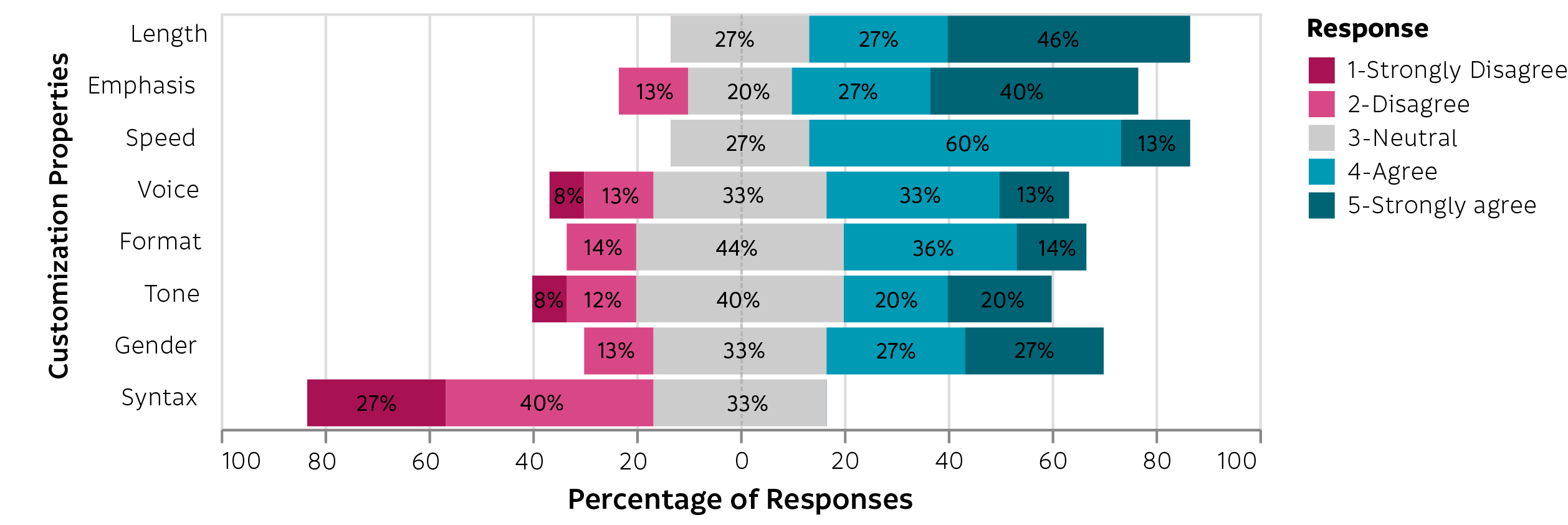}
        \caption{Summary of Likert scale questionnaire responses on customization properties preference by participants. Participants rated how much they agreed that the customization property could help them to consume AD more effectively}
    \Description[Diverging bar chart for Customization Preference Likert scale Rating]{This is a diverging bar chart that summarises the percentage of responses for each customization property. The y-axis indicates different customization properties, which are, from bottom to top, syntax, gender, tone, format, voice, speed, emphasis, and length. The x-axis, bounded between 0 to 100 to the left and 0 to 100 to the right, indicates the percentages of Likert scale responses from the participants. Each bar consists of multiple sub-bars in which the size is proportional to the percentage of response. Each sub-bar shows data for different Likert scale responses.
    }
    \label{fig:customization_preference}
\end{figure*}

Participants generally responded positively to the customization properties presented in the study. 
Most participants strongly agreed that customizing \textit{length} ($Mean = 4.20; SD = 0.86$) and \textit{emphasis} ($Mean = 3.93; SD = 1.10$) helped consume AD more effectively. 
They agreed that adjusting the AD speed was also desirable ($Mean = 3.87; SD = 0.64$). 
They were neutral about \textit{voice} ($Mean = 3.33; SD = 1.11$), \textit{format} ($Mean = 3.47; SD = 0.92$), \textit{tone} ($Mean =3.33; SD = 1.18$), and \textit{gender} ($Mean = 3.67; SD = 1.05$).
\textit{Syntax} customization was the only property that they perceived not beneficial. 
Fig.~\ref{fig:customization_preference} summarizes the Likert-scale responses for each customization property.

In an open discussion, three participants noted they would prefer not to customize ADs due to concerns about potential complexities associated with interacting with the setting user interface. 
For instance, P3, who self-identified as less tech-savvy, preferred minimal interaction or even no customization despite recognizing the potential advantages of it. 

\subsection{Participants’ Preferences on Customization Properties}

\subsubsection{Length (Succinct, Verbose, Very Verbose)}
Most participants (N = 13) stated that they would appreciate the ability to customize the length of ADs, because it would allow them to tailor the amount of information to their individual preferences and satisfy their curiosity about a scene. 

\vspace{-0.2pc}

\begin{quote}
    \textit{``I like it because, you know, different people have different preferences and curiosity. So some people might want to know more visual information, some people might want to know less.''} - P1
\end{quote}
  
Six participants shared that the need to adjust AD length depended on the type of video and their intention to watch it. 
For example, participants expected more detailed AD to fully follow the instructions in instructional videos that required complete understanding. 
P15 said, \textit{``It depends on the intention. For example, if it's like an exercise video, then I will adjust it to be very descriptive because maybe I want to know what's the accurate form [...]. But then if, for example, it's like a music video or something similar, then probably I wouldn't need a very detailed description.''} (P15) 

\subsubsection{Emphasis (Person, Activity, Object, Setting)}
Nine participants believed that customizing the ADs' information \textit{emphasis} could enhance the clarity and focus of the AD. 
The option was deemed particularly helpful as different videos have various intended learning outcomes, and participants may want to focus on different aspects of the content.

\begin{quote}
    \textit{``I think [emphasis customization is] good because everyone have a very different appetite or interest. Let's say for example, I'm going to be watching an instructional video. Then I would want more information on the activities. [...] So, that would help [to follow the instructions]. And then if, let's say, I am watching a documentary, maybe I like to have [AD] to focus more on the people’s descriptions.''} - P10
\end{quote}

On the other hand, three participants were concerned about lacking the knowledge to decide what \textit{emphasis} property setting to choose. Thus, they suggested encouraging the creators to determine the information focus when creating the AD. 

\begin{quote}
   \textit{It's very difficult [to decide] the good setting for [emphasis] customization, especially for people who have not watched the video, you don't even know what you want to focus on [...].}– P8
\end{quote}

\subsubsection{Speed (Slow, Fast, Original)}

Participants (N=12) mentioned that adjusting the speed enhanced the audio clarity of the video. 
For example, when the AD was too fast, the participant could slow it down, which helps to digest the AD better.

Also, the adjustment of AD speed was mentioned by five participants as a potential enhancement to information navigation efficiency while watching videos. 
It is based on their familiarity with consuming audio content at a faster pace. 
Increasing the speed enabled participants to quickly grasp the main message and navigate to the desired information in the AD.

Two participants reported that increasing the AD speed could potentially enhance their enjoyment of videos. 
In particular, when AD was presented in a synthesizer voice by default, which tended to be more robotic and monotonous compared to the human voice. 
The faster speed helped to maintain their interest and prevent them from losing focus.

\subsubsection{Voice (Human, Synthesizer)}
Nine participants preferred human voice over synthesizer voice and consistently chose the human voice option whenever possible. 
For this reason, voice customization was deemed not useful. 
P14 said, \textit{``I think [...] anybody will always prefer a human voice because it makes the whole thing sound natural.''}

But, some participants pointed out the advantages of using a synthesizer. 
For instance, P8 noted that the synthesizer voice could be easier to adjust the pitch or tone without sounding unnatural. 
Additionally, P4 mentioned that the synthesizer can be easier to comprehend sometimes than the human voice due to the variability in human accents.

\subsubsection{Format (Inline, Extended)}
Nine participants expressed concerns that extended descriptions negatively impacted their viewing experience because they disliked the idea of videos being paused abruptly. For instance, P4 said, \textit{``Sometimes you don't want to cut into the flow of actual video itself. So that the [audio description] can be played seamlessly.''} Nevertheless, the participants would be open to an extended format if it added value, such as providing more detailed descriptions during pause. Two participants were more positive; they  mentioned that adjusting the format can enhance the clarity of the AD. Specifically, the extended version of AD allowed the participants to have more time to digest the information and understand it better. 

Format customization can be beneficial in different ways depending on the type of video (N=6). For example, P2 found that extended versions were particularly helpful for videos where the dialogue or monologue was unrelated to the visual content being displayed. However, in fast-paced videos like movies, participants preferred inline descriptions.

\begin{quote}
    \textit{``It really depends on the kind of video that you're playing. Okay, for example, some videos, that's movies they're very fast moving you will need something in-line. I mean you cannot pause it and then let the [audio descriptions play]. It will break the flow and I will not enjoy that.''} - P3
\end{quote}

\subsubsection{Tone (Monotonous, Dynamic)}
Participants believed that customizing the tone to fit into the overall video’s mood (\textit{i.e.}, dynamic setting) would enhance the immersion (N=9). P1 said, \textit{``Audio descriptions can be sounded in a more “annoyed” tone to really emphasize that the user is annoyed with the situation. You know, it then conveys more emotion in the audio descriptions.''}

Other than conveying emotion, the different tones of AD can make the video more engaging and less monotonous. \textit{``Maybe [in] a love story, when the couple is happy, the describer can speak in a more upbeat tone. Then let's say if they break [up], then the describer described in a very sad tone.''} - P11

\subsubsection{Gender (Female, Male)}
Participants were mainly indifferent to gender voice customization as long as there was enough contrast between the AD and the dialogue or monologue narrator (N=6). Aligned with \cite{netflixADGuide}, regardless of the gender voice, the AD should contrast with most of the voices in the videos.

\subsubsection{Syntax (Present Tense, Past Tense)}
All but one did not find syntax customization to be meaningful. They believed that customizing the syntax would not have any impact on consuming AD. 
However, one participant mentioned that having the option to present the AD in different syntax helps her to feel the temporal factor of the video, that is to indicate whether things are happening in the present or the past.

\subsubsection{Emergent customization property}
While we designed our study to focus on eight customization properties (\textit{i.e.}, six presentation and two content customization properties), additional customization of interest emerged. 
Three participants recommended allowing users to customize the \textit{language} of AD. 
Currently, AD in online streaming platforms is typically provided in the same language as the video's language. 
While these platforms may offer dubbing options for the video, the AD remains in the original language. 
P15 said, \textit{``Maybe language. Because sometimes I want to watch Korean dramas, but because there's no dubbed [audio descriptions], then I can't watch it.'' } (P15).

\subsection{Section Summary}
Overall, participants expressed that customizing properties like \textit{length, emphasis, speed, voice, format, tone,} and \textit{language} would positively impact video understanding, audio clarity, immersion, and information navigation efficiency. Also, customization would allow participants to have the flexibility to personalize their experience based on individual preferences. However, some concerns were raised about potential interruptions caused by specific customization properties, such as format. Moreover, there is a need to have intuitive usability and reasonable defaults for those who are not tech-savvy or prefer minimal interaction and customization with the system. Lastly, participants expressed uncertainty about selecting desirable settings, e.g., determining the specific information to focus on for the \textit{emphasis} property.

\begin{figure*}
    \centering
    \includegraphics[width=16cm]{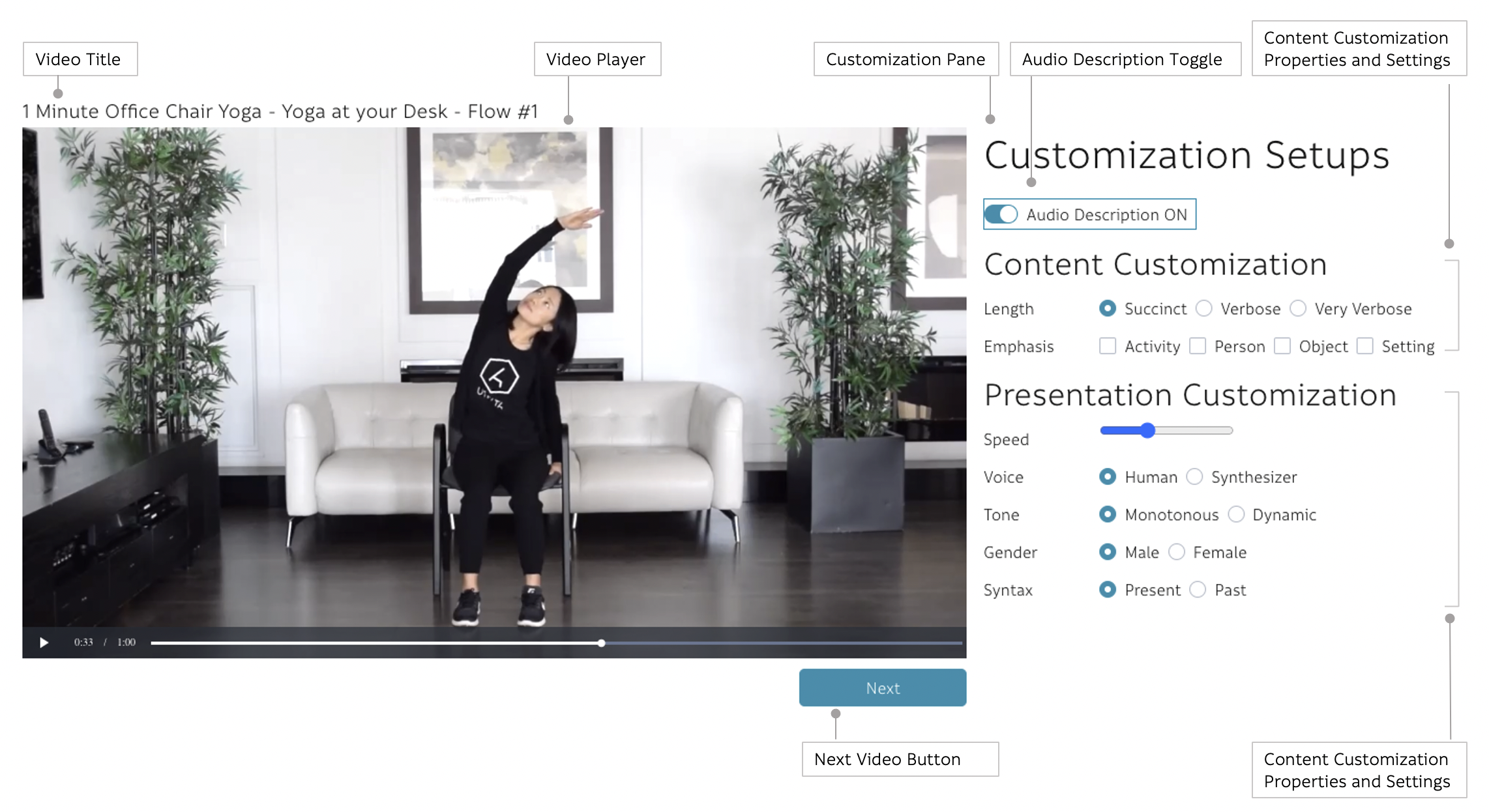}
        \caption{CustomAD interface consists of a video player (left), which allows users to play, pause, and seek the video, and a customization pane (right) where users can customize the properties of ADs. The customization properties are grouped into content settings and presentation settings. The content customization adjusts the script's content as users change the ADs length and emphasis. In presentation customization, users could adjust speed, voice, tone, gender, and grammatical syntax of the ADs to change how the ADs are read out. Users can also toggle the ADs on and off.}
    \Description[Screenshot of the interface.]{
    Figure of the interface screenshots. The interface is divided into two horizontally, left and right components. The left component shows the video player and the right component shows the customization setups.
    }
    \label{fig:prototype_figure}
\end{figure*}

\section{CustomAD: A Hi-fi Prototype for Audio Description Customization}\label{sec:customad}

Insights from Study 1 shed light on BLV participants' desire for AD customization, its perceived benefit, and the potential trade-offs tied to different customization properties.  
These observations has motivated us to conduct an experiment to substantiate utility of AD customization through a controlled study. 
As a groundwork to conduct the experiment, we designed and developed the high-fidelity web prototype named CustomAD. 

CustomAD is a prototype designed to support AD customization for BLV individuals. 
The system consists of two main parts: a video pane on the left and a customization pane on the right (Fig. \ref{fig:prototype_figure}). 
Users can play, pause, and seek the video with the video pane. 
On the right side of the interface, there is the customization setups pane, which displays various customization properties that users can modify to tailor their experience. 
The interface can also be fully operated with keyboard shortcuts for accessibility (Appendix 1). 

Using CustomAD, participants can customize seven properties: \textit{length, emphasis, speed, voice, tone, gender,} and \textit{syntax} (Table~\ref{tab:customAD_customization_properties}). 
These customization properties are grouped into two categories: content and presentation. 
Content customizations focus on the \textit{length} and \textit{emphasis} properties, while presentation customizations involve \textit{speed, voice, tone, gender}, and \textit{syntax}. 
For each property, we offer two to four settings to choose from, except for the \textit{speed} property, where the user adjusts the value using a slider;
the participant can adjust \textit{speed} property from 0.25x to 2x in increments of 0.25 using the slider. This setting choice was inspired from existing video streaming platform like, YouTube, which allowed a speed range between 0.25x to 2x using slider.
For \textit{length}, the settings are \textit{succinct}, \textit{verbose}, and \textit{very verbose}. 
For \textit{emphasis}, participants can choose to focus the ADs on descriptions of \textit{activity}, \textit{person}, \textit{object}, or \textit{setting}. 
For the \textit{voice} property, the options are \textit{human} or \textit{synthesizer}. 
For \textit{tone}, the options are \textit{monotonous} or \textit{dynamic}.
For \textit{gender}, the options are \textit{male} or \textit{female}. 
And for \textit{syntax}, the choices are between \textit{present} and \textit{past}. 
The default settings for \textit{length, emphasis, speed, voice, tone, gender,} and \textit{syntax} are \textit{succinct}, \textit{null}, \textit{1.0x}, \textit{human}, \textit{monotonous}, \textit{male}, and \textit{present}, respectively. 
Any changes in customization settings take immediate effect on ADs. 

The findings from Study 1 informed how we chose the list of customizable properties and how we designed CustomAD.
For instance, both \textit{length} and \textit{format} properties are made customizable by manipulating the \textit{length} option; as a user adjusts the option between \textit{succinct}, \textit{verbose}, and \textit{very verbose}, the system automatically switches the AD format between normal and extended AD format. This design decision was appropriate because most participants would want to have an extended AD that pauses a video to fit an AD only when the long description added substantially more information.
 
We applied speed customization across the entire video, not just the AD segment to minimize unnecessary silence when the participant is speeding up the AD. 
Though the majority considered syntax customization unnecessary, we kept this property to address possible long-tail user requirements. 
The videos we used in Study 2 that we describe below were in English, and all participants were fluent in English. 
Thus, we left the evaluation the \textit{language} property that surfaced through Study 1 for future research. 
We gave the participants the option of not to set a value to \textit{emphasis} property and leave it empty for a more balanced emphasis on the information. 
This decision was also made to address the participants' concern about not knowing what setting they should choose in information emphasis.

\section{Study 2: Evaluation Method}\label{sec:summative_study_method}
To investigate the effectiveness of AD customization for video consumption and to evaluate how BLV individuals perceive its impact, we conducted remote user study with 12 BLV people. 
 
The study is a two (\textit{with-} and \textit{without-customization}) by three (\textit{entertainment}, \textit{explainer}, \textit{tutorial} videos) within-subjects design. 
We chose video type as our independent variable because videos’ visual content and AD presentation varied depending on the video’s intended audience, goal, and tone, which in turn could influence the utility of customization. 
To mitigate potential learning effect, we counterbalanced the sequence in which participants engaged with video customization and encountered different video types. 

\subsection{Videos}\label{sec:summative_videos}
We used three different videos in our study---\textit{entertainment}, \textit{explainer}, and \textit{tutorial}. 
We selected these video types because of their popularity among online video viewers \cite{googleInsight,pictory}. 
Although the list provided is not specifically for BLV individuals, we believe that what is commonly watched by sighted people should also be accessible for BLV individuals. 
Beyond their popularity, these three video types also capture the diversity of the videos' delivery styles, objectives, and ways of consumption which is suitable to evaluate the impact of customization on different video types. 
To see variability within the type of videos, we selected two videos to watch for each type. 
We chose videos that are about one- to two-minute because it is suitable for the duration of each study session.
The videos we used are as follows (see the example scenes for each video in Fig.~\ref{fig:summative_videos}):

\begin{figure*}
    \centering
    \includegraphics[width=16cm]{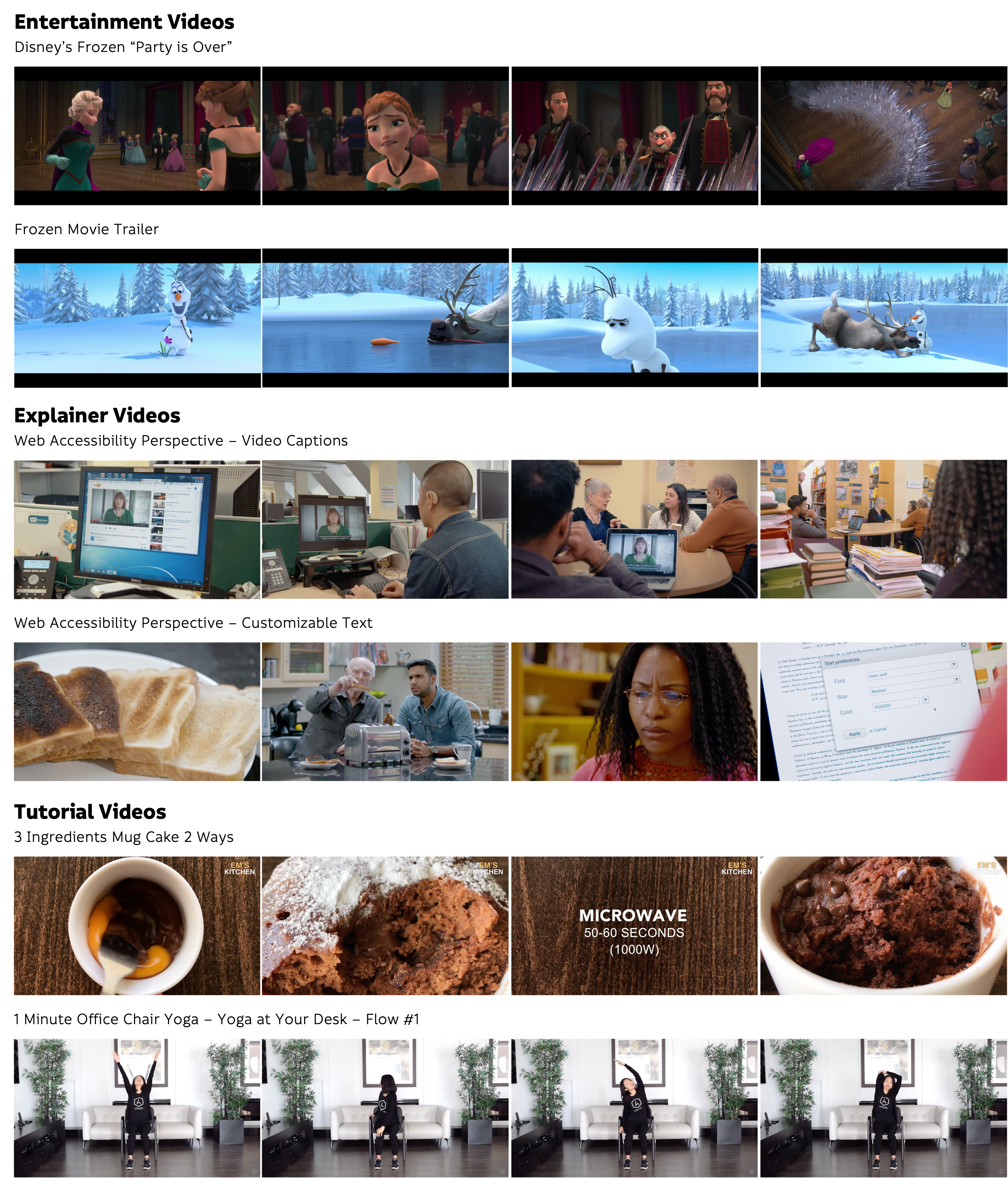}
        \caption{Videos used in Study 2. We used six videos of three types: \textit{entertainment}, \textit{explainer}, and \textit{tutorial}. Each row represents a video. }
    \Description[Screenshots of scenes in the evaluation study's videos]{
    Screenshots of several scenes in the videos we used in our evaluation study. The figure is presented in grid format. Each row represents one video type, from top to bottom, that is music, instructional, entertainment video, campaign, explainer, advertisement, and documentary videos. We show four example scenes for each video.
    }
\label{fig:summative_videos}
\end{figure*}

\begin{itemize}
    \item Entertainment Video 1 (EN1): Disney’s Frozen ``Party is Over''~\cite{en1}, Duration: 49 seconds. A short clip of Disney’s Frozen that shows two main characters arguing in a ball reception.
    \item Entertainment Video 2 (EN2): Frozen Movie Trailer, Duration~\cite{en2}: 1 min 30 sec. A short animation video that mainly shows the characters, Olaf, a snowman, and Sven, and a reindeer fighting over a carrot, which appears to be Olaf’s nose in the snowy open area. 
    \item Explainer Video 1 (EX1): Web Accessibility Perspective – Video Captions~\cite{ex1}, Duration: 1 min 17 seconds. This video explains the importance of video captions. The visuals in the videos illustrate how video captions are used in different circumstances, in which the visuals also complement the audio explanations. 
    \item Explainer Video 2 (EX2): Web Accessibility Perspective – Customizable Text~\cite{ex2}, Duration: 1 min 17 seconds. This video explains the importance of having the ability to customize texts in different interfaces. The visuals in the videos illustrate how customization is useful for catering to different user needs and abilities.
    \item Tutorial Video (Tut1): 3 Ingredients Nutella Mug Cake 2 Ways~\cite{tut1}, Duration: 2 mins. This video shows a series of instructions for making two versions of Nutella cakes that are made in mugs. The video shows a top-down view of a pair of hands performing the cooking, all the ingredients, and the tools for cooking. Also, the video shows the text of the ingredient measurements.
    \item Tutorial Video (Tut2): 1 Minute Office Chair Yoga – Yoga at Your Desk – Flow \#1~\cite{tut2}, Duration: 1 min. This video shows a woman who is sitting on the chair she usually uses for work, demonstrating several yoga movements that can be done just by sitting on her office chair.
\end{itemize}

\subsection{Diverse AD Versions for Customization}
For each video, we worked with an expert audio describer to create a diverse set of AD versions. 
The expert has received professional training (\textit{i.e.,} Bonnie Barlow Creative Audio and Visual Describing) on audio descriptions and worked on multiple projects of creating ADs for over two years now. 
The expert, being a native English speaker, is particularly suitable for our projects as all our videos are in English. 
The expert generated fifteen distinct AD scripts for each video (see Appendix 2 for all the AD combinations scripts). 
These encompassed variations driven by the \textit{length} property (\textit{i.e.}, \textit{succinct}, \textit{verbose}, \textit{very verbose}), as well as combinations of the \textit{length} and \textit{emphasis} properties ($3$ \textit{length} $\times$ $4$ \textit{emphasis} $= 12$ combinations, \textit{e.g.,} Succinct $\times$ Activity, Verbose $\times$ Person, and so on). For several videos, such as EN1, EX1, and EX2, were originally accompanied by the succinct version of AD. As a result, for these specific videos, the expert describers created only the remaining versions of AD. 

After receiving the AD scripts from the experts, we generate different versions of AD for Presentation Customization with DupDub.\footnote{\url{https://www.dupdub.com/}} Using this website, we are able to generate different voices, style, gender, and tone which is particularly useful voice, tone, and gender properties.

\subsection{Participants}\label{sec:summative_participants}
We recruited twelve BLV participants who were familiar with AD (six males and six females, aged between 24 and 62; Mean = 37.92, SD = 12.34). 
Seven participants’ impairments were congenital, while five participants acquired their impairments later in their lives. 
There are five total blind and seven low-vision participants. 
Ten participants have also participated in the Study 1 (see Table~\ref{tab:participant_demographic}).

\subsection{Metrics}
To assess the benefits and drawbacks of AD customization, we evaluated several metrics fundamental to effective video consumption. 
We measured participants’ video understanding, time cost and prototype usability, cognitive load, and perceived usefulness, particularly in clarity, immersion, and information navigation efficiency. 
The primary goal of watching videos is often to actively seek information to gain new knowledge, followed closely by entertainment and enjoyment \cite{googleInsight}. 
Thus, it was important to evaluate how customization plays a role in video understanding and immersion. 
We also evaluated task completion time and frequency of customization settings changes to better understand the user's interaction pattern with AD customization and its time cost. 
Also, we evaluated prototype usability, task cognitive load, and perceived usefulness as our Study 1 revealed they were also the main factor that affects the overall experience with AD customization. 

We used an information-seeking task as the proxy to gauge video understanding. 
We curated prompts that covered dimensions like person, activities, settings, emotions, video purpose, and narrative, aimed to stimulate the information-seeking tasks. These dimensions were based on essential visual description elements in videos \cite{stangl2020person}. Please refer to Appendix 3 for all the prompts. We also collected the user interaction log that consists of the changes in customization and timestamp information. We used the log data to calculate task completion time and customization modification frequency. We also evaluate the system's usability using the SUS questionnaire \cite{jordan1996usability} and exit interviews. We measured cognitive demand via the NASA Task Load Index (TLX) \cite{colligan2015cognitive}, using a 7-point Likert scale, like in previous studies \cite{huh2023avscript,huh2023genassist}. Additionally, participants completed Likert scale questionnaires and an exit interview to determine customization's perceived utility across video types. They considered the statement: \textit{``Customization in [video types] enhances [clarity/immersion/information navigation efficiency] of audio descriptions.''} We further discuss the participant’s rationale for each rating in exit interviews.

\subsection{Procedure}
At the start of the user study, we provided participants with a detailed explanation of the research's aims, the concept of customizing AD, and the upcoming tasks they will perform in the session. Also, we got their verbal consent to participate in the online study. We thoroughly explained the functionalities and the keyboard shortcuts available within the CustomAD system. Subsequently, participants used CustomAD to watch a total of six videos---two from each distinct video type (\textit{entertainment}, \textit{explainer}, and \textit{tutorial} videos). Participants could do AD customizations for three videos, while for the other three videos, they were not able to do AD customizations. To familiarize participants with the CustomAD interface, we introduced a practice video which is an explainer video of color contrast by W3C\footnote{\url{https://www.youtube.com/watch?v=a9kNUv6N8Rk&ab_channel=W3CWebAccessibilityInitiative\%28WAI\%29}}. This facilitated their engagement and familiarity with the prototype until they felt confident to progress to the primary task.

After the participants were familiar with CustomAD, they started the information-seeking task, followed by answering the NASA TLX questionnaire after each video, and ended with subjective ratings of perceived usefulness and exit interview. We opted for information-seeking tasks as a proxy for video understanding because information-seeking is often the primary motivation behind watching videos and to gain an understanding of the video to absorb new knowledge \cite{googleInsight}. In addition, we believed that answering these questions would also encourage the participants to perform AD customization, something they might have not been familiar with yet. The prompts were given before the participants started watching the video and the NASA TLX questionnaires were administered after every video. After participants were done with the six videos, participants completed the System Usability Scale (SUS) questionnaire and the subjective ratings, along with the exit semi-structured interview. The study lasted for 2 hours. To ensure thorough analysis, we recorded and transcribed screen and audio interactions. We also recorded their interaction log at the backend, which consists of customization properties, setting changes, and timestamp information. Our study was approved by our institution’s IRB, and participants consented to participate in the study through verbal consent before the study started. Participants were compensated with \$40 for their participation.

\section{Study 2: Evaluation Result}
We adopted a mixed-methods approach and performed both quantitative and qualitative analyses of our data. We collected the video and Zoom screen recording, the answers to the information-seeking prompts, the survey responses to perform both quantitative and qualitative analyses, and the interaction log with the system. We reviewed the proportion of the questions in which the participants got the correct answer over the total number of questions. We also analyzed the NASA TLX task-load questionnaire Likert-scale results. We reviewed both the session recording and interaction logs to extract participants’ interaction with- and without- customization conditions. We transcribed the exit interviews and grouped them according to (1) the perceived usefulness of the customization in general, also in different video types and (2) the usability of the system that supports the customization.

\subsection{Correctness of Information Seeking Prompts}

\begin{figure}
    \centering
    \includegraphics[width=7.5cm]{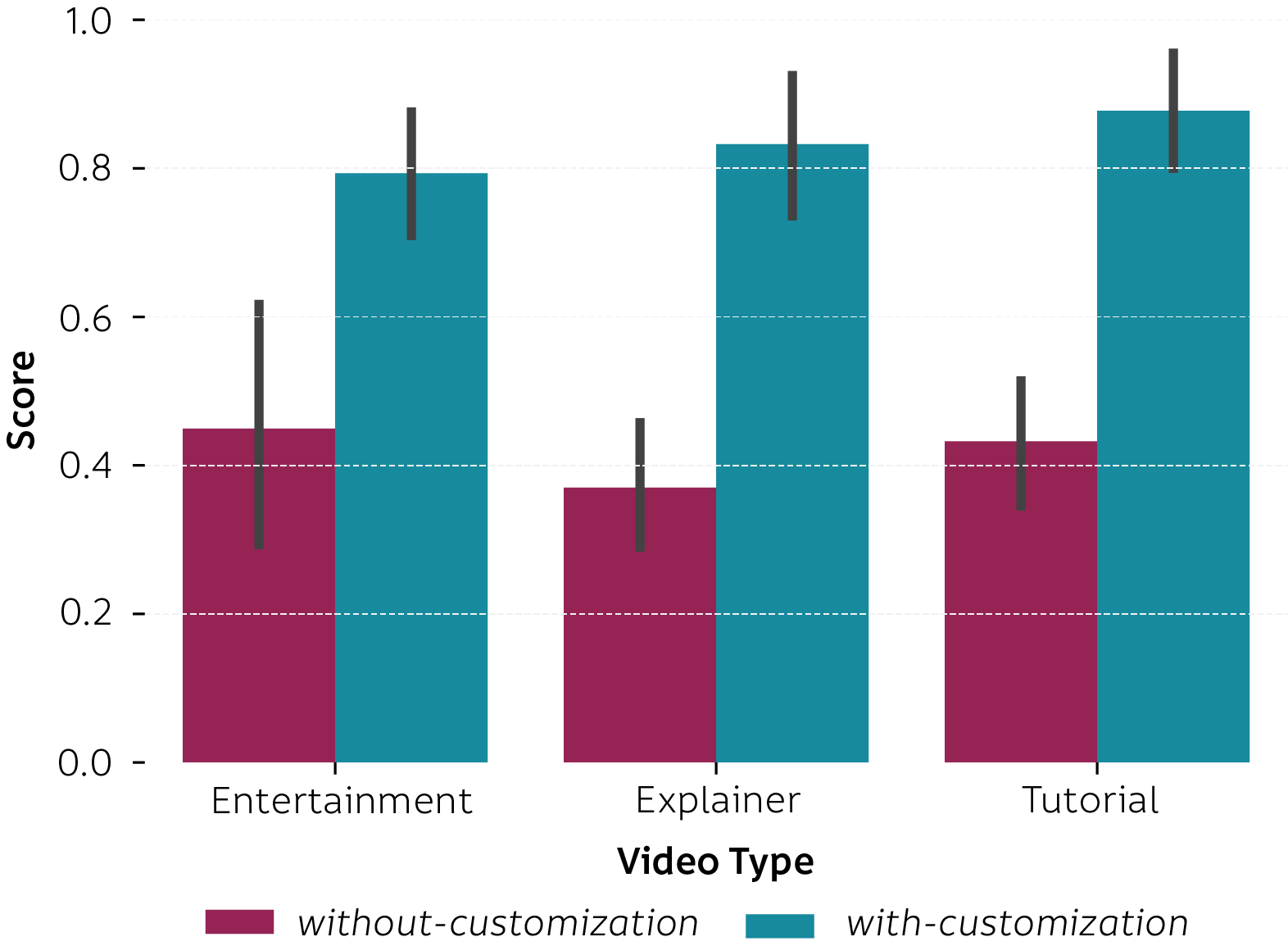}
        \caption{Average correctness scores for completing information seeking tasks for  different video types (\textit{entertainment}, \textit{explainer}, \textit{tutorial}) and interface conditions (\textit{without-customization} and \textit{with-customization}). The vertical line on each bar represents standard deviation.}
    \Description[A bar chart for the correctness of prompts]{A bar chart that summarises the average value of the correctness of prompts. The x-axis indicates the video types and the y-axis indicates the correctness scores. Each video type has two bar visualizations that show the without-customization condition and the with-customization condition. Each bar also shows the error bar.
    }
    \vspace{-1pc}
    \label{fig:correctness}
\end{figure}

Participants’ performance in answering the prompts was significantly higher in the \textit{with-customization} condition (Fig~\ref{fig:correctness}). We obtained the score by calculating the proportion of the questions in which the participants got the correct answer over the total number of questions per video. On average, scores were 0.83 (SD = 0.16) with customization, compared to 0.42 (SD = 0.21) without customization. This trend extended across different video types. For \textit{entertainment}, \textit{explainer}, and \textit{tutorial} videos, participants using customization achieved averages of 0.79, 0.83, and 0.88, respectively (SD = 0.15, 0.17, 0.14). 
In contrast, scores for participants when consuming videos without customization were 0.45, 0.37, and 0.43, respectively (SD = 0.29, 0.16, 0.16). 
A Generalized Linear Mixed-Model (GLMM) analysis with binomial distribution family and logit link function, where the customization and the video type are the fixed effect and participant was the random effect, showed the difference in accuracy ($z = 4.185$; $p < 0.001$) was significant. 
GLMMs were particularly suited for regression analyses involving dependent variables bounded between 0 and 1. Additionally, GLMMs can effectively include the random effect to account for subject-specific variability for \textit{within-subject} study design.
This result suggests that customization enabled participants to tailor their AD, resulting in enhanced prompt accuracy, which indicated better video understanding.

\subsection{Time and Interaction Analysis}
Using the interaction log data, we calculated the duration participants took to watch the video and answer information-seeking prompts. 
To cater to diverse videos' original duration, we normalized the value to duration per video minute. 
Table~\ref{tab:duration} offers a detailed breakdown of task completion time, presented in video minute units. 
The \textit{with-customization} condition took longer compared to the \textit{without-customization} condition. Specifically, under the \textit{with-customization} scenario, participants averaged 11.03 minutes per video minute (SD = 6.98, Md = 9.39). 
Conversely, the \textit{without-customization} condition recorded an average of 6.69 minutes per video minute (SD = 6.98). 
The extended duration in the \textit{with-customization} condition was anticipated as it is attributed to the time invested in the customization process. 
In addition, customizable properties such as \textit{length} and \textit{speed} properties also extended video duration. 

\begin{table}[]
\begin{tabular}{p{2.9cm}llll}
\hline
          &               & \multicolumn{3}{c}{\makecell{Task Completion Time \\ (per video minute)}} \\ \hline
Condition & Video Types   & Mean         & SD        & Md.        \\ \hline
\textit{without-customization} & \textit{Overall} & 6.69  & 4.23 & 5.60 \\
\textit{} & Entertainment & 6.52         & 4.04                  & 6.25          \\
\textit{} & Explainer     & 6.38         & 4.89                  & 5.16          \\
\textit{} & Tutorial      & 7.18         & 4.05                  & 6.79          \\ \hline
\textit{with-customization}    & \textit{Overall} & 11.03 & 6.98 & 9.39 \\
\textit{} & Entertainment & 13.93        & 10.56                 & 12.91         \\
\textit{} & Explainer     & 9.28         & 3.45                  & 8.96          \\
\textit{} & Tutorial      & 9.87         & 4.19                  & 8.98          \\ \hline
\end{tabular}
\caption{Summary of task completion time for \textit{without-customization} and \textit{with-customization} conditions. We present the overall task completion time along with the duration for \textit{entertainment}, \textit{explainer}, and \textit{tutorial} videos. We have normalized the values to duration per video minute (pvm) as durations varied across videos.}
\vspace{-2pc}
\label{tab:duration}
\end{table}

In our evaluation of task completion time across different video types, distinct patterns emerged when comparing conditions \textit{with-} and \textit{without-customization}. With customization enabled, participants took the longest on Entertainment videos, on average, at 13.93 minutes (SD = 10.56 minutes), followed by Tutorial videos at 9.28 minutes (SD = 3.45 minutes), and finally explainer videos at 9.87 minutes (SD = 4.19 minutes). In contrast, without customization, the pattern shifted: Participants took the longest to complete tutorial videos, which was 7.18 minutes on average (SD = 4.05 minutes), followed by entertainment at 6.52 minutes (SD = 4.04 minutes), and explainer videos, which was 6.38 minutes, on average (SD = 4.89 minutes). A GLMM analysis with gamma distribution family and log link function, where the customization and video types were the fixed effect and participant was the random effect showed a significant difference in completion time between the \textit{with-} and \textit{without-customization} conditions ($z = 5.733; p < 0.001$). 

From the interaction log, we also assessed the frequency of various customizations. The \textit{emphasis} property was modified most frequently (N = 91), followed by \textit{length} (N = 59), \textit{speed} (N = 35), \textit{tone} (N = 18), \textit{gender} (N = 11), \textit{syntax} (N = 11), and \textit{voice} (N = 4). On average, customization patterns among participants were as follows: 8 for \textit{emphasis}, 5 for \textit{length}, 3 for \textit{speed}, 2 for \textit{tone}, and 1 each for \textit{gender} and \textit{syntax}. Participants performed almost no voice customization. These trends mirror insights from our Study 1. Participants had previously expressed a strong preference for \textit{length} and \textit{emphasis} customizations, emphasizing their value for enhancing video understanding. This was evidenced by their leading customization counts in Study 2. Participants held neutral opinions on tone and gender customizations, as these settings were infrequently adjusted in Study 2. Interestingly, though syntax customization was deemed less useful in the Study 1, some participants (N = 5) still opted for it in the Study 2. A closer examination of the data revealed a predominant preference for the ``Present'' syntax setting (N\textsubscript{present} = 8, N\textsubscript{past} = 3). As for the voice property, despite some customization considerations performed by participants, the manual review indicated a dominant selection of the ``Human'' voice, matching the preference identified in our Study 1.

During the exit interview, seven participants noted that the extended duration of the videos due to customization might be the greatest drawback to the overall experience. However, they also emphasized that the benefits of customization, such as gaining more visual information to understand the video’s content and making the video more engaging, outweighed the additional duration required. For example, P15 mentioned:

\begin{quote}
    \textit{"With customization, watching videos indeed have become longer, but if that helps me to know more detail of the scene, why not?"} - P15
\end{quote}

\subsection{Usability}
CustomAD's usability in the \textit{with-customization} condition obtained a notable SUS score of 84.23 (SD = 11.92), placing it in the `Excellent' usability rating. From our semi-structured interviews, a notable portion of respondents (N=10) emphasized the value of keyboard shortcuts for enhanced ease of use. 

\begin{quote}
   \textit{"I like the fact that I can use keyboard shortcut, it makes navigation between customization properties simple and fast."} - P6 
\end{quote}

Additionally, participants (N=2) appreciated the streamlined, linear navigation and minimalist design, which was tailored and sufficient for task completion and improved video and AD consumption. 

\begin{quote}
   \textit{"From the keyboard shortcut and the voice navigation, I can feel that the system is very simple, clean, and easy to navigate. I think this is very important as customization itself is very complex, so easy navigation between customization properties is very crucial."} - P11
\end{quote}

Another notable feature was the instant reflection of customization changes, allowing participants to immediately perceive modifications without unnecessary delay. However, there was feedback from one participant regarding potential disturbance when the menu voice synthesizer played concurrently with the video. Potential solutions suggested included the use of sound beeps in lieu of full-sentence customization readings or employing audio ducking—reducing the background or video volume when vocalizing customization commands. In line with evolving viewing habits, three participants underscored the relevance of extending CustomAD's compatibility to mobile devices, emphasizing their growing preference for mobile-based video consumption over traditional platforms like laptops or TVs.

\subsection{Task Load}
\begin{figure*}
    \centering
    \includegraphics[width=16cm]{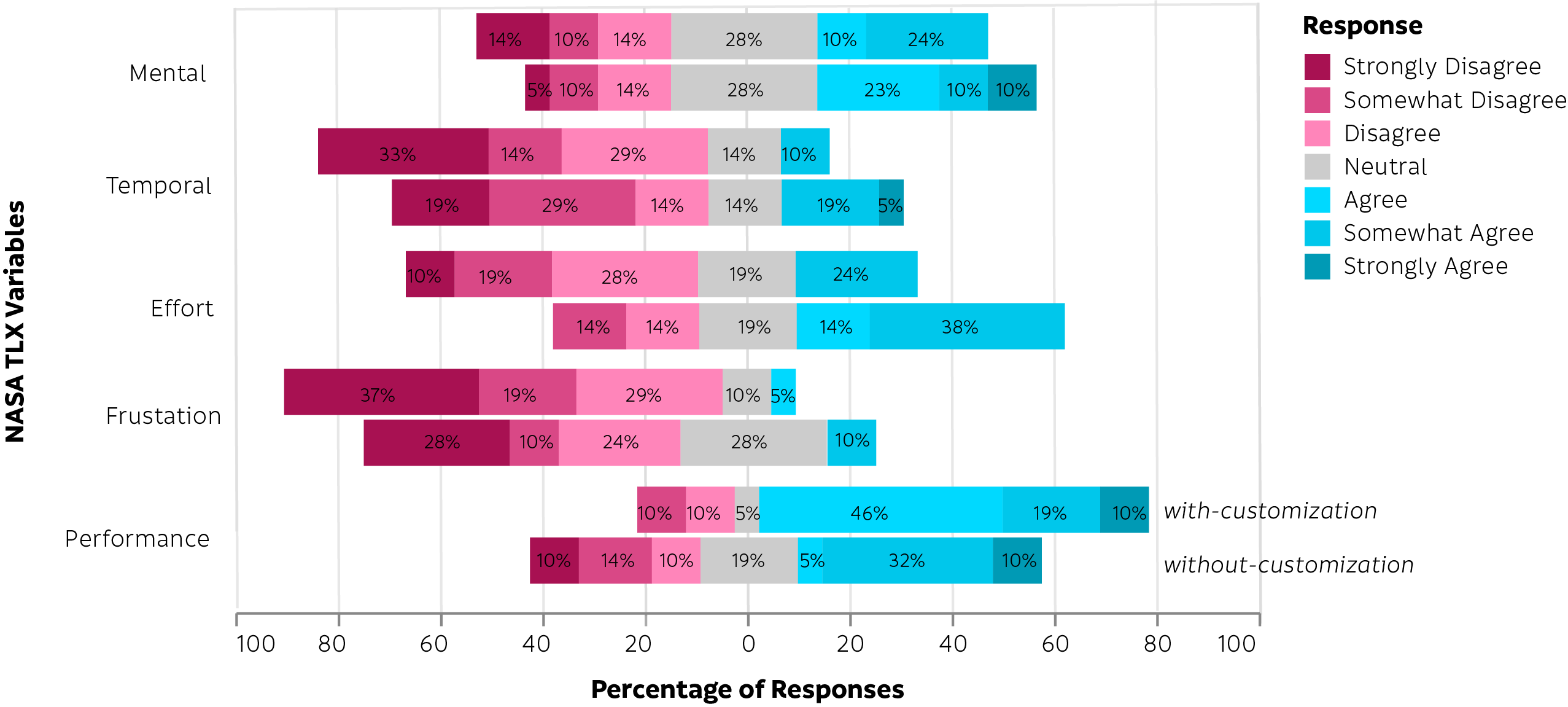}
    \vspace{-1pc}
        \caption{Summary of Likert-scale response for each NASA TLX variable, which are mental, temporal, effort, frustration, and performance. For mental, temporal, effort, and frustration, lower scores are better, while for performance higher the better.}
    \Description[Diverging bar chart for the task cognitive load analysis.]{
    This is a diverging bar chart that summarises the percentage of responses for each NASA TLX variable. The x-axis, bounded between 0 to 100 to the left and 0 to 100 to the right, indicates the percentages of Likert scale responses from the participants. The y-axis indicates the different NASA TLX variables. Each bar consists of multiple sub-bars in which the size is proportional to the percentage of response. Each sub-bar shows data for different Likert scale responses. We also visualize the summary for two different study conditions, which are with-customization (on the top) and without-customization (on the bottom).
    }
\label{fig:tlx}
\end{figure*}

Overall, participants rated the task load as lower when using customization than without customization while achieving high perceived performance (Fig.~\ref{fig:tlx}). We report the mean and standard deviation in tuple with the following format: Mean = (\textit{with-}, \textit{without-customization}), SD = (\textit{with-}, \textit{without-customization}). Specifically, participants reported lower mental load (Mean = (3.67, 4.38); SD = (1.56, 1.69)), temporal load (Mean = (2.52, 3.05); SD = (1.69, 1.36)), effort load (Mean = (3.29, 4.24), SD = (1.31, 1.30), and frustration load (Mean = (2.29, 2.89), SD = (1.35, 1.40)) with customization compared to without customization. Moreover, participants also perceived a higher performance when using customization to answer prompts (Mean = (5.14, 4.05), SD = (1.49, 1.75)). We omitted the physical effort variable in NASA TLX, as the task didn't involve any physical activity. A statistical analysis with Wilcoxon Signed-Rank test showed a significant difference in both effort ($W = 20; p = 0.04 < 0.05$) and performance  ($W = 30; p = 0.01 < 0.05$). 

Our exit interview results also supported this. None of the participants found the customization process, particularly for information-seeking tasks, to be difficult or excessively demanding. Considering the benefits they received from customization, participants were willing to tolerate the additional effort required.

\begin{quote}
    \textit{"If you talk about effort, of course, i need to put more compare to watching the videos without customization. I think this is given. But, i think it is still bearable, i still feel the overall task load is minimal, considering the additional benefit, which is to enjoy and understand the video more, I would get from putting a slight more effort."} - P6  
\end{quote}

These results underscore the positive impact of customization. Interestingly, while the Study 1 highlighted concerns about increased effort and potential troubles with customization, our results indicate the opposite. Participants' reports of reduced task load suggest a favorable potential for adopting customization. The consistently low task load scores highlight the benefit of AD customization.

\subsection{Perceived Usefulness of Customization}
\begin{figure*}
    \centering
    \includegraphics[width=16cm]{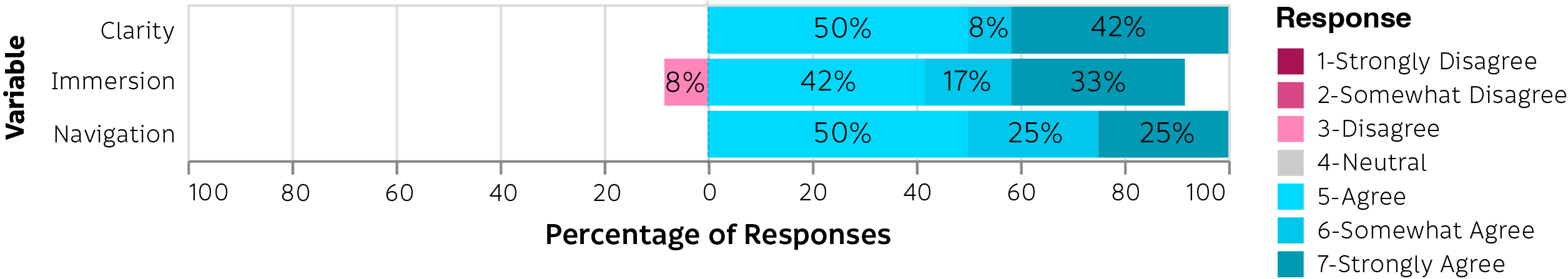}
        \caption{Summary of Likert-scale responses of perceived usefulness of AD customization for enhancing AD clarity, immersion, and information navigation efficiency.}
    \Description[Diverging bar chart for perceived usefulness of audio descriptions.]{This is diverging bar chart which summarises the percentage of responses for each variable for perceived usefulness of AD customization. The x-axis, bounded between 0 to 100 to the left and 0 to 100 to the right, indicates the percentages of Likert scale responses from the participants. The y-axis indicates the different perceived usefulness variables. Each bar consists of multiples sub-bars in which the size is proportional to the percentage of response. Each sub-bar shows data for different Likert scale responses.
    }
    \vspace{-1pc}
    \label{fig:usefulness}
\end{figure*}

In terms of the perceived usefulness of customization, participants agreed that being able to customize contributed to better clarity, immersion, and information navigation efficiency of watching the video. Closer inspection at Fig.~\ref{fig:usefulness} reveals the rating scale is 6.33, 5.92, 6 for clarity, immersion, and information navigation efficiency, respectively (clarity: SD = 0.65; immersion: SD = 1.16; information navigation: SD = 0.74). This result suggests a broad consensus: AD customization demonstrably improved the video’s clarity, immersion, and information navigation efficiency.

Delving deeper into the clarity, participants found advantages from the flexibility of information consumption enabled by the customization. The capacity to select pertinent information, also the ability to customize to an individual’s information preference and requirements, helped the participants to consume the AD more effectively. 

Participants mentioned the positive impact of customization on video immersion, though two cited interruptions as a drawback. The flexibility of AD length and emphasis emerged as particularly valuable, ensuring a nuanced capture of entire emotions and ambiance. For example, adjustments to the \textit{emphasis} property, especially choosing the ``Setting'' setting enhanced the understanding of the video's narrative and ambiance better than staying with the default. The exit interviews revealed that customization options such as \textit{emphasis}, \textit{length}, and \textit{tone} were particularly beneficial for participants with residual vision (N = 2). For instance, P16 mentioned that \textit{emphasis} customization allowed him to supplement his limited visual input and foster a more integrative viewing experience by obtaining information about the scene’s mood.

\begin{quote}
    \textit{``with some residual [vision], I can still somehow see visuals related to object and actions, sometimes, but maybe, not so much on the overall mood. Then, I decided to choose “Setting” [in emphasis property], which I think, then, I can obtain a better understanding of the mood, whether it is happy or sad.''} - P16 
\end{quote}

Though customization is perceived beneficial for good immersion of the video, however, certain customizations, such as adjusting the \textit{length} property, require the videos to pause to accommodate extended AD. This customization makes participants feel uncomfortable because the videos’ flow is interrupted most of the time. These findings are consistent with insights from the Study 1.

The flexibility of customizing the content (\textit{length} and \textit{emphasis} properties) and presentation, especially change in \textit{speed} property, is deemed beneficial in increasing information navigation efficiency. Our study required the participants to answer prompts and participants (N=5) found that customizing \textit{length} and \textit{emphasis} quickly helped them to obtain the information they needed. For example, P12 mentioned:

\begin{quote}
    \textit{``I think emphasis property is very useful to quickly get the answer I know. While answering the questions, I was kind of deciding which category of information this question belongs to, for example, if the question is about a human or the person, then I quickly change the emphasis to “Person”, then I can quickly get the answer I need. Putting the task aside, in general, I like the fact I can customize the emphasis of information. With that, I can decide what information I want to mostly be informed about, perhaps, it also depends on the video I am watching.''} - P12
\end{quote}

In addition to \textit{emphasis}, \textit{speed} was also perceived to be useful to increase efficiency (N = 3). Particularly, it is related to how quick or slow a user can obtain the necessary information in video content. For example, P1, who typically preferred to listen to videos at a slower speed, used to watch videos he found too fast multiple times to understand the content. However, by allowing him to adjust the speed, he could choose a pace that he felt comfortable with and could follow along more easily.

\begin{quote}
    \textit{``I like the ability to adjust the speed. For example, when some videos have very fast audio descriptions, or the video speed is generally too fast, I have to watch the video multiple times to understand the video, but being able to adjust the speed, allows me to choose which speed is more comfortable for me and I can follow, then I don’t need to watch the video multiple times''} - P1
\end{quote}

\section{Discussion and Future Work}
Our research aims to enhance the video-watching experience for Blind and Low Vision (BLV) individuals by enabling them to customize Audio Descriptions (ADs). A formative study involving BLV individuals revealed a strong preference for AD customization, highlighting key properties such as \textit{length, emphasis, speed, voice, format, tone} and \textit{language}. Building on these insights, we then sought conclusive evidence on the impact of customization on the BLV video-watching experience. To this end, we developed CustomAD, a high-fidelity web prototype that supported AD customization. CustomAD let BLV individuals to tailor both the content and presentation of ADs, including properties such as \textit{length, information emphasis, speed, voice, tone, gender, } and \textit{syntax}. The evaluation results indicated that CustomAD significantly enhanced the video-watching experience for BLV individuals. The tool deepened comprehension of visual content, increased immersion, and improved efficiency in information navigation.

We discuss the trade-offs of customization and task load and duration of videos. While our Study 1 revealed some participants' concerns regarding potential interruptions and cognitive load, Study 2 alleviated this concern. Our analysis of NASA TLX results showed that participants felt AD customization imposed minimal cognitive strain. This minimal load might also be attributed to the system's high usability, as indicated by its ``excellent'' rating on the SUS, which likely reduced perceived interruptions and mental demand. Moreover, the benefit of doing customization (\textit{e.g.,} obtaining more detailed information) seems to compensate for the added complexity and possible extended video duration, making customization still favorable. 

\subsection{Interactions for Adjusting Customization Options}
In CustomAD's current design, BLV individuals manually adjust settings to align with their preferences. This autonomy to customize the AD, while offering flexibility to match the ADs to what they prefer, has surfaced challenges; people lacked the contextual or objective knowledge needed to pinpoint appropriate settings. This challenge was particularly pronounced for properties like \textit{emphasis} and \textit{tone}, where the appropriate setting was influenced by the video's context and learning objectives---factors users may not know in advance. Such feedback underscores the need for offering users good defaults and setting recommendations based on a video’s context or learning objectives. For example, the system could suggest popular settings chosen by prior viewers, thereby aiding subsequent viewers in their customization choices. In addition, video creators should also play a crucial role in better communicating the intended objective of the videos and assign good AD defaults. Beyond human-centric solutions, there is a compelling opportunity to leverage automation. The system could, for example, analyze a video's content and automatically suggest relevant \textit{emphasis} settings. Additionally, leveraging a video’s sentiment and topic, \textit{e.g.,} \cite{yadav2015multimodal,wu2022video} could inform the preferable AD tone, guiding users between a monotonous or dynamic delivery.

\subsection{Generating Multiple Versions of AD}
In our study, we opted for professional audio describers to generate diverse versions of AD to explore customization properties. While this approach produced high-quality AD, it induced significant time and financial costs. For instance, the creation of a minute of video content required 15 distinct scripts (\textit{i.e.}, three different AD lengths and the combinations of 3 \textit{length} $\times$ 4 \textit{emphasis} properties), costing around \$60 in total per video minute. Moreover, the turnaround time took more than a week, especially for visually complex videos. Given these constraints, there is an apparent need to explore more economical and efficient avenues. Notably, research by Natalie et al. \cite{natalie2023supporting} underscores the viability of novice describers as a cost-effective alternative, although they often face challenges balancing quality dimensions such as succinctness versus sufficiency. An intriguing future direction might involve coupling novice-generated verbose ADs with Large Language Models (LLM) and perhaps also computer vision (CV) models for automatic refinement. For example, LLM could help in summarizing tasks to produce AD in different verbosity levels (\textit{length} property). Also, LLM could help in rephrasing AD that is specific to the description of ``Activity'', ``Person'', ``Object'', and ``Setting'' (\textit{emphasis} property) depending on the primary focus of the video detected by CV models. Leveraging this human-AI collaboration method could pave the way to address the identified challenges and signify a promising avenue in AD research. 
Furthermore, the prospect of fully automating AD generation for diverse customization properties is a compelling research trajectory. The growing field of intelligent visual descriptions in recent studies [\textit{e.g.,} \cite{lei2020mart,wang2021toward,radford2021learning}] indicates progress, though there is more to be done to fully achieve optimal quality in AD. 

\subsection{AD Customization in Co-watching Activities}
In our findings, customization emerged as an enabler for enhancing video comprehension and immersion, allowing individuals to tailor AD to their preferences. The present study's focus has been predominantly on the effect of customization for solo-viewing scenarios. However, this perspective may not directly translate to co-viewing experiences, where diverse preferences diverge, and individual adjustments can become a point of disagreement. When multiple viewers with distinct preferences watch content concurrently, an intriguing challenge arises: \textit{How might a system balance and accommodate these distinct preferences without prioritizing one viewer's experience over another?} Potential solutions could involve collaborative customization, where participants input their preferences, and an intelligent system aggregates and weighs them to derive a consensus-based customization setting. Or, allowing the use of their individual headset, but still ensuring interactivity to support co-watching experience. However, as AD customization remains a nascent field with limited empirical exploration, our study's insights into its effects on solo-viewing remain relevant. Future endeavors in this domain would benefit from delving into collaborative customization paradigms to enhance co-viewing experiences.

\subsection{Toward More Accessible Videos Players and Streaming Platforms}
We suggested existing video players and online streaming platforms to incorporate AD customization features on top of toggling AD on and off. Our research highlights both the desire and the benefits of AD customization, which many current platforms overlook. Looking ahead, we envision evolving CustomAD into a plug-in compatible with popular video players like VLC and QuickTime Player, enhancing their AD delivery capabilities. Feedback from participants further underscores the importance of supporting AD customization for mobile devices. This is because most of them are consuming videos on mobile devices daily. In addition, given that approximately 4.18 billion individuals consumed video content on mobile devices in 2020 — a number that is expected to double annually \cite{yansmedia} — optimizing CustomAD for diverse devices is essential. Moreover, online streaming platforms such as YouTube, Netflix, and YouDescribe could enhance accessibility for blind audiences by incorporating these customization features. Most video streaming platforms possess existing customization features, like speed, language, and video quality adjustments, and showed that users are already familiar with them. Extending beyond these customization features and implementation, we envision extending customization features to include AD customization features, enriching the viewing experience for BLV individuals.

\subsection{Broadening Customization to Other Audio-Visual Based Technologies}
Our study, while focused on traditional videos, highlights potential applications of customization for assisting BLV individuals across various visual-audio-based technologies. The core principles shared by conventional videos with AD and emerging technologies—reliant on visual and audio explanations for accessibility, such as AR/VR ~\cite{chang2022omniscribe, lo2021navigation, heap2023experiencing}, remote sighted assistance (e.g., Be My Eyes~\cite{BeMyEyes}, Seeing AI~\cite{SeeingAI}, Be Specular~\cite{BeSpecular}, and navigational tools (e.g.,~\cite{Soundscape})—indicate a potential for applying AD customization in these technologies. For example, in the context of 360° videos, often used for BLV entertainment~\cite{chang2022omniscribe}, education~\cite{heap2023experiencing}, or navigation~\cite{lo2021navigation} purposes, customization could allow BLV users to tailor information emphasis and verbosity. Such customization capabilities may enable BLV individuals to filter out irrelevant information and focus on areas of interest, addressing the immersion and cognitive load trade-off in a full 360° viewpoint~\cite{chang2022omniscribe}. Similarly, customization may benefit remote-sighted assistance technologies, where users can specify details like information emphasis and length for clearer and more focused assistance to understand the surroundings. The customization capabilities can also extend to the presentation of instructions, including adjustable properties such as speed, voice, and gender. Moreover, these technologies should consider expanding customization options to include properties such as viewing direction and the frequency of descriptions, which would be based on the distance at which the descriptions are triggered and read out.. Additionally, for audio-based navigation context, the ability to adjust instruction audio speed and length may significantly improve real-world mobility for BLV individuals. Customization adaptation not only preserves the essence of ADs in traditional video contexts but also broaden its utility to enhance world interaction for the BLV community.

\vspace{-0.1cm}

\section{Limitations}
Study 2 assessed the value of customization for information-seeking tasks rather than pure content comprehension. This design choice was deliberate; we aimed to simulate scenarios of information seeking and to actively prompt participants towards customization. While this was essential in designing the study, our approach might deviate from the typical viewing behavior of a BLV individual. Future research could delve deeper, examining additional effectiveness, benefits and trade-offs of customization. 

Our study was conducted in a controlled environment, limiting our ability to assess the effectiveness of customization in more naturalistic settings. Furthermore, the videos used in the study were relatively short. Although participants engaged with six videos for an average of eleven minutes each, totaling around an hour of exposure to CustomAD, we acknowledged that studying CustomAD usage with longer videos could provide additional insights into how customization settings persist or evolve over time.

Field studies conducted in more organic, longitudinal settings — where BLV individuals customize their viewing experiences without prompts — might shed the light the nuanced interplay between customization, video comprehension, and viewing experience. Additionally, exploring community-driven and automated methods for dynamic customization in longer videos would be a valuable avenue for future research.

\section{Conclusion}
In this paper, we explored the potential benefits and challenges of audio description (AD) customization through an interview and an evaluation of a high-fidelity prototype for AD customization, CustomAD. The results of the interview study with 15 BLV participants highlighted the perceived benefits of customization, such as enhanced clarity and understanding of video visuals, improved immersion, and increased information navigation efficiency. Despite these advantages, participants expressed concerns regarding possible interruptions and the challenges of determining appropriate customization settings due to a lack of prior knowledge of the video's context and learning objectives. Recognizing these trade-offs, we designed and developed a high-fidelity prototype, called CustomAD, which supports customization of \textit{length, emphasis, speed, voice, tone, gender}, and \textit{syntax} properties. We conducted an evaluation using CustomAD with 12 BLV participants. This evaluation affirmed that customization empowers BLV users to enhance their understanding of videos, experience greater immersion, and incur minimal mental load. We concluded the study by discussing key insights and suggesting enhancements for CustomAD and similar AD customization platforms. In particular, we see the potential in leveraging both crowdsourcing and automation to streamline the process of producing customized AD versions, informing the customization setting selection, effect on the co-watching experience, and integrating customization in existing video players and online. 

\begin{acks}
This research / project is supported by the Ministry of Education, Singapore under its Academic Research Fund Tier 2 (Project ID: T2EP20220-0016). Any opinions, findings and conclusions or recommendations expressed in this material are those of the author(s) and do not reflect the views of the Ministry of Education, Singapore.
\end{acks}

\bibliographystyle{ACM-Reference-Format}
\bibliography{customad}


\begin{thebibliography}{108}


\ifx \showCODEN    \undefined \def \showCODEN     #1{\unskip}     \fi
\ifx \showDOI      \undefined \def \showDOI       #1{#1}\fi
\ifx \showISBNx    \undefined \def \showISBNx     #1{\unskip}     \fi
\ifx \showISBNxiii \undefined \def \showISBNxiii  #1{\unskip}     \fi
\ifx \showISSN     \undefined \def \showISSN      #1{\unskip}     \fi
\ifx \showLCCN     \undefined \def \showLCCN      #1{\unskip}     \fi
\ifx \shownote     \undefined \def \shownote      #1{#1}          \fi
\ifx \showarticletitle \undefined \def \showarticletitle #1{#1}   \fi
\ifx \showURL      \undefined \def \showURL       {\relax}        \fi
\providecommand\bibfield[2]{#2}
\providecommand\bibinfo[2]{#2}
\providecommand\natexlab[1]{#1}
\providecommand\showeprint[2][]{arXiv:#2}

\bibitem[3PlayMedia(2020)]%
        {3PlayMedia_BeginnerGuideline}
\bibfield{author}{\bibinfo{person}{3PlayMedia}.} \bibinfo{year}{2020}\natexlab{}.
\newblock \bibinfo{title}{Beginner's Guide to Audio Description}.
\newblock \bibinfo{howpublished}{\url{https://go.3playmedia.com/hubfs/WP\%20PDFs/Beginners-Guide-to-Audio-Description.pdf}}.
\newblock
\newblock
\shownote{Accessed: 2021-01-13}.


\bibitem[Abla et~al\mbox{.}(2010)]%
        {abla2010customizable}
\bibfield{author}{\bibinfo{person}{G. Abla}, \bibinfo{person}{E.N. Kim}, \bibinfo{person}{D.P. Schissel}, {and} \bibinfo{person}{S.M. Flanagan}.} \bibinfo{year}{2010}\natexlab{}.
\newblock \showarticletitle{Customizable scientific web portal for fusion research}.
\newblock \bibinfo{journal}{\emph{Fusion Engineering and Design}} \bibinfo{volume}{85}, \bibinfo{number}{3} (\bibinfo{year}{2010}), \bibinfo{pages}{603--607}.
\newblock
\showISSN{0920-3796}
\urldef\tempurl%
\url{https://doi.org/10.1016/j.fusengdes.2010.02.030}
\showDOI{\tempurl}
\newblock
\shownote{Proceedings of the 7th IAEA Technical Meeting on Control, Data Acquisition, and Remote Participation for Fusion Research}.


\bibitem[Access(2023)]%
        {NVDA}
\bibfield{author}{\bibinfo{person}{NV Access}.} \bibinfo{year}{2023}\natexlab{}.
\newblock \bibinfo{title}{NVDA Version 2023.3.4}.
\newblock \bibinfo{howpublished}{\url{https://www.nvaccess.org/download/}}.
\newblock
\newblock
\shownote{Accessed: 2023-09-14}.


\bibitem[ADLab(2024)]%
        {adlabGuideline}
\bibfield{author}{\bibinfo{person}{ADLab}.} \bibinfo{year}{2024}\natexlab{}.
\newblock \bibinfo{title}{ADLab Audio Descriptioon: Lifeling Access for the Blind}.
\newblock \bibinfo{howpublished}{\url{http://www.adlabproject.eu/Docs/adlab\%20book/}}.
\newblock
\newblock
\shownote{Accessed: 2024-03-14}.


\bibitem[AI(2024)]%
        {SeeingAI}
\bibfield{author}{\bibinfo{person}{Seeing AI}.} \bibinfo{year}{2024}\natexlab{}.
\newblock \bibinfo{title}{Seeing AI: Talking Camera for the Blind}.
\newblock \bibinfo{howpublished}{\url{https://www.seeingai.com/}}.
\newblock
\newblock
\shownote{Accessed: 2024-09-24}.


\bibitem[Amazon(2023)]%
        {amazonPrime}
\bibfield{author}{\bibinfo{person}{Amazon}.} \bibinfo{year}{2023}\natexlab{}.
\newblock \bibinfo{title}{Amazon Prime}.
\newblock \bibinfo{howpublished}{\url{http://www.amazon.com}}.
\newblock
\newblock
\shownote{Accessed: 2023-09-14}.


\bibitem[American Council of~the Blind(2017)]%
        {acb_guideline}
\bibfield{author}{\bibinfo{person}{Audio Description~Project American Council of~the Blind}.} \bibinfo{year}{2017}\natexlab{}.
\newblock \bibinfo{title}{Guideline for Audio Describers}.
\newblock \bibinfo{howpublished}{\url{https://www.acb.org/adp/guidelines.html}}.
\newblock
\newblock
\shownote{Accessed: 2020-11-6}.


\bibitem[Apple(2023)]%
        {VoiceOver}
\bibfield{author}{\bibinfo{person}{Apple}.} \bibinfo{year}{2023}\natexlab{}.
\newblock \bibinfo{title}{Chapter 1. Introducing VoiceOver}.
\newblock \bibinfo{howpublished}{\url{https://www.apple.com/voiceover/info/guide/_1121.html}}.
\newblock
\newblock
\shownote{Accessed: 2023-09-14}.


\bibitem[Australia(2015)]%
        {campaignAustralia}
\bibfield{author}{\bibinfo{person}{Vision Australia}.} \bibinfo{year}{2015}\natexlab{}.
\newblock \bibinfo{title}{Audio Description on TV}.
\newblock \bibinfo{howpublished}{\url{https://youtu.be/ULgLZn91TMo}}.
\newblock
\newblock
\shownote{Accessed: 2023-11-6}.


\bibitem[Beam and Kosicki(2014)]%
        {beam2014personalized}
\bibfield{author}{\bibinfo{person}{Michael~A. Beam} {and} \bibinfo{person}{Gerald~M. Kosicki}.} \bibinfo{year}{2014}\natexlab{}.
\newblock \showarticletitle{Personalized News Portals: Filtering Systems and Increased News Exposure}.
\newblock \bibinfo{journal}{\emph{Journalism \& Mass Communication Quarterly}} \bibinfo{volume}{91}, \bibinfo{number}{1} (\bibinfo{year}{2014}), \bibinfo{pages}{59--77}.
\newblock
\urldef\tempurl%
\url{https://doi.org/10.1177/1077699013514411}
\showDOI{\tempurl}
\showeprint{https://doi.org/10.1177/1077699013514411}


\bibitem[Bineeth~Kuriakose and Sandnes(2022)]%
        {kuriakose2022tools}
\bibfield{author}{\bibinfo{person}{Raju~Shrestha Bineeth~Kuriakose} {and} \bibinfo{person}{Frode~Eika Sandnes}.} \bibinfo{year}{2022}\natexlab{}.
\newblock \showarticletitle{Tools and Technologies for Blind and Visually Impaired Navigation Support: A Review}.
\newblock \bibinfo{journal}{\emph{IETE Technical Review}} \bibinfo{volume}{39}, \bibinfo{number}{1} (\bibinfo{year}{2022}), \bibinfo{pages}{3--18}.
\newblock
\urldef\tempurl%
\url{https://doi.org/10.1080/02564602.2020.1819893}
\showDOI{\tempurl}
\showeprint{https://doi.org/10.1080/02564602.2020.1819893}


\bibitem[Blind Citizen~Australia(2024)]%
        {BlindCitizenAustralia}
\bibfield{author}{\bibinfo{person}{Australian Communications Consumer Action Network Media Access~Australia Blind Citizen~Australia, Vision~Australia}.} \bibinfo{year}{2024}\natexlab{}.
\newblock \bibinfo{title}{Blindness Sector Report on the 2012 ABC Audio Description Trial}.
\newblock \bibinfo{howpublished}{\url{https://www.bca.org.au/wp- content/uploads/2018/04/Blindness_Sector_Report_on_the_2012_ABC_Audio_Description_Trial.doc}}.
\newblock
\newblock
\shownote{Accessed: 2024-03-14}.


\bibitem[Bragg et~al\mbox{.}(2021)]%
        {bragg2021expanding}
\bibfield{author}{\bibinfo{person}{Danielle Bragg}, \bibinfo{person}{Katharina Reinecke}, {and} \bibinfo{person}{Richard~E. Ladner}.} \bibinfo{year}{2021}\natexlab{}.
\newblock \showarticletitle{Expanding a Large Inclusive Study of Human Listening Rates}.
\newblock \bibinfo{journal}{\emph{ACM Trans. Access. Comput.}} \bibinfo{volume}{14}, \bibinfo{number}{3}, Article \bibinfo{articleno}{12} (\bibinfo{date}{jul} \bibinfo{year}{2021}), \bibinfo{numpages}{26}~pages.
\newblock
\showISSN{1936-7228}
\urldef\tempurl%
\url{https://doi.org/10.1145/3461700}
\showDOI{\tempurl}


\bibitem[Broadcasting(2023)]%
        {ndr}
\bibfield{author}{\bibinfo{person}{Northern~German Broadcasting}.} \bibinfo{year}{2023}\natexlab{}.
\newblock \bibinfo{title}{Audio description guidelines}.
\newblock \bibinfo{howpublished}{\url{https://www.ndr.de/fernsehen/barrierefreie_angebote/audiodeskription/Audio-description-guidelines,audiodeskription142.html}}.
\newblock
\newblock
\shownote{Accessed: 2023-04-09}.


\bibitem[Bunt et~al\mbox{.}(2007)]%
        {bunt2007supporting}
\bibfield{author}{\bibinfo{person}{Andrea Bunt}, \bibinfo{person}{Cristina Conati}, {and} \bibinfo{person}{Joanna McGrenere}.} \bibinfo{year}{2007}\natexlab{}.
\newblock \showarticletitle{Supporting interface customization using a mixed-initiative approach}. In \bibinfo{booktitle}{\emph{Proceedings of the 12th International Conference on Intelligent User Interfaces}} (Honolulu, Hawaii, USA) \emph{(\bibinfo{series}{IUI '07})}. \bibinfo{publisher}{Association for Computing Machinery}, \bibinfo{address}{New York, NY, USA}, \bibinfo{pages}{92–101}.
\newblock
\showISBNx{1595934812}
\urldef\tempurl%
\url{https://doi.org/10.1145/1216295.1216317}
\showDOI{\tempurl}


\bibitem[Caldwell et~al\mbox{.}(2008)]%
        {wcag2008}
\bibfield{author}{\bibinfo{person}{Ben Caldwell}, \bibinfo{person}{Michael Cooper}, \bibinfo{person}{Loretta~Guarino Reid}, \bibinfo{person}{Gregg Vanderheiden}, \bibinfo{person}{Wendy Chisholm}, \bibinfo{person}{John Slatin}, {and} \bibinfo{person}{Jason White}.} \bibinfo{year}{2008}\natexlab{}.
\newblock \showarticletitle{Web content accessibility guidelines (WCAG) 2.0}.
\newblock \bibinfo{journal}{\emph{WWW Consortium (W3C)}} (\bibinfo{year}{2008}).
\newblock


\bibitem[Campos et~al\mbox{.}(2020)]%
        {campos2020cinead}
\bibfield{author}{\bibinfo{person}{Virginia~P Campos}, \bibinfo{person}{Tiago~MU de Ara{\'u}jo}, \bibinfo{person}{Guido~L de Souza~Filho}, {and} \bibinfo{person}{Luiz~MG Gon{\c{c}}alves}.} \bibinfo{year}{2020}\natexlab{}.
\newblock \showarticletitle{CineAD: a system for automated audio description script generation for the visually impaired}.
\newblock \bibinfo{journal}{\emph{Universal Access in the Information Society}} \bibinfo{volume}{19}, \bibinfo{number}{1} (\bibinfo{year}{2020}), \bibinfo{pages}{99--111}.
\newblock
\urldef\tempurl%
\url{https://doi.org/10.1007/s10209-018-0634-4}
\showDOI{\tempurl}


\bibitem[Canada(2023)]%
        {mac}
\bibfield{author}{\bibinfo{person}{Media~Access Canada}.} \bibinfo{year}{2023}\natexlab{}.
\newblock \bibinfo{title}{DESCRIPTIVE VIDEO PRODUCTION AND PRESENTATION BEST PRACTICES GUIDE FOR DIGITAL ENVIRONMENTS}.
\newblock \bibinfo{howpublished}{\url{http://www.mediac.ca/DVBPGDE_V2_28Feb2012.asp}}.
\newblock
\newblock
\shownote{Accessed: 2023-04-09}.


\bibitem[CBM(2023)]%
        {mariaDocument}
\bibfield{author}{\bibinfo{person}{EndTheCycle CBM}.} \bibinfo{year}{2023}\natexlab{}.
\newblock \bibinfo{title}{My Story: Maria (with Extended Audio Description)}.
\newblock \bibinfo{howpublished}{\url{https://youtu.be/JUIJ_aNxsG8}}.
\newblock
\newblock
\shownote{Accessed: 2023-11-6}.


\bibitem[Chang et~al\mbox{.}(2022)]%
        {chang2022omniscribe}
\bibfield{author}{\bibinfo{person}{Ruei-Che Chang}, \bibinfo{person}{Chao-Hsien Ting}, \bibinfo{person}{Chia-Sheng Hung}, \bibinfo{person}{Wan-Chen Lee}, \bibinfo{person}{Liang-Jin Chen}, \bibinfo{person}{Yu-Tzu Chao}, \bibinfo{person}{Bing-Yu Chen}, {and} \bibinfo{person}{Anhong Guo}.} \bibinfo{year}{2022}\natexlab{}.
\newblock \showarticletitle{OmniScribe: Authoring Immersive Audio Descriptions for 360° Videos}. In \bibinfo{booktitle}{\emph{Proceedings of the 35th Annual ACM Symposium on User Interface Software and Technology}} (Bend, OR, USA) \emph{(\bibinfo{series}{UIST '22})}. \bibinfo{publisher}{Association for Computing Machinery}, \bibinfo{address}{New York, NY, USA}, Article \bibinfo{articleno}{15}, \bibinfo{numpages}{14}~pages.
\newblock
\showISBNx{9781450393201}
\urldef\tempurl%
\url{https://doi.org/10.1145/3526113.3545613}
\showDOI{\tempurl}


\bibitem[Chmiel and Mazur(2016)]%
        {chmiel2016researching}
\bibfield{author}{\bibinfo{person}{Agnieszka Chmiel} {and} \bibinfo{person}{Iwona Mazur}.} \bibinfo{year}{2016}\natexlab{}.
\newblock \showarticletitle{Researching preferences of audio description users—Limitations and solutions}.
\newblock \bibinfo{journal}{\emph{Across Languages and Cultures}} \bibinfo{volume}{17}, \bibinfo{number}{2} (\bibinfo{year}{2016}), \bibinfo{pages}{271--288}.
\newblock
\urldef\tempurl%
\url{https://doi.org/10.1556/084.2016.17.2.7}
\showDOI{\tempurl}


\bibitem[Chmiel and Mazur(2022)]%
        {chmiel2022homogenous}
\bibfield{author}{\bibinfo{person}{Agnieszka Chmiel} {and} \bibinfo{person}{Iwona Mazur}.} \bibinfo{year}{2022}\natexlab{}.
\newblock \showarticletitle{A homogenous or heterogeneous audience? Audio description preferences of persons with congenital blindness, non-congenital blindness and low vision}.
\newblock \bibinfo{journal}{\emph{Perspectives}} \bibinfo{volume}{30}, \bibinfo{number}{3} (\bibinfo{year}{2022}), \bibinfo{pages}{552--567}.
\newblock
\urldef\tempurl%
\url{https://doi.org/10.1080/0907676X.2021.1913198}
\showDOI{\tempurl}
\showeprint{https://doi.org/10.1080/0907676X.2021.1913198}


\bibitem[Colligan et~al\mbox{.}(2015)]%
        {colligan2015cognitive}
\bibfield{author}{\bibinfo{person}{Lacey Colligan}, \bibinfo{person}{Henry~WW Potts}, \bibinfo{person}{Chelsea~T Finn}, {and} \bibinfo{person}{Robert~A Sinkin}.} \bibinfo{year}{2015}\natexlab{}.
\newblock \showarticletitle{Cognitive workload changes for nurses transitioning from a legacy system with paper documentation to a commercial electronic health record}.
\newblock \bibinfo{journal}{\emph{International journal of medical informatics}} \bibinfo{volume}{84}, \bibinfo{number}{7} (\bibinfo{year}{2015}), \bibinfo{pages}{469--476}.
\newblock
\urldef\tempurl%
\url{https://doi.org/10.1016/j.ijmedinf.2015.03.003}
\showDOI{\tempurl}


\bibitem[Commission(2020)]%
        {fcc2020}
\bibfield{author}{\bibinfo{person}{Federal~Communications Commission}.} \bibinfo{year}{2020}\natexlab{}.
\newblock \bibinfo{title}{21st Century Communications and Video Accessibility Act (CVAA)}.
\newblock \bibinfo{howpublished}{\url{https://www.fcc.gov/consumers/guides/21st-century-communications-and-video-accessibility-act-cvaa}}.
\newblock
\newblock
\shownote{Accessed: 2020-11-6}.


\bibitem[Cronin and King(1990)]%
        {cronin1990development}
\bibfield{author}{\bibinfo{person}{Barry~J Cronin} {and} \bibinfo{person}{Sharon~Robertson King}.} \bibinfo{year}{1990}\natexlab{}.
\newblock \showarticletitle{The Development of the Descriptive Video Servicesm}.
\newblock \bibinfo{journal}{\emph{Journal of Visual Impairment \& Blindness}} \bibinfo{volume}{84}, \bibinfo{number}{10} (\bibinfo{year}{1990}), \bibinfo{pages}{503--506}.
\newblock


\bibitem[Damian et~al\mbox{.}(2011)]%
        {damian2011individualized}
\bibfield{author}{\bibinfo{person}{Ionut Damian}, \bibinfo{person}{Birgit Endrass}, \bibinfo{person}{Peter Huber}, \bibinfo{person}{Nikolaus Bee}, {and} \bibinfo{person}{Elisabeth Andr{\'e}}.} \bibinfo{year}{2011}\natexlab{}.
\newblock \showarticletitle{Individualized Agent Interactions}. In \bibinfo{booktitle}{\emph{Motion in Games}}, \bibfield{editor}{\bibinfo{person}{Jan~M. Allbeck} {and} \bibinfo{person}{Petros Faloutsos}} (Eds.). \bibinfo{publisher}{Springer Berlin Heidelberg}, \bibinfo{address}{Berlin, Heidelberg}, \bibinfo{pages}{15--26}.
\newblock
\showISBNx{978-3-642-25090-3}


\bibitem[de~Santana et~al\mbox{.}(2013)]%
        {de2013firefixia}
\bibfield{author}{\bibinfo{person}{Vagner~Figueredo de Santana}, \bibinfo{person}{Rosimeire de Oliveira}, \bibinfo{person}{Leonelo Dell~Anhol Almeida}, {and} \bibinfo{person}{Marcia Ito}.} \bibinfo{year}{2013}\natexlab{}.
\newblock \showarticletitle{Firefixia: an accessibility web browser customization toolbar for people with dyslexia}. In \bibinfo{booktitle}{\emph{Proceedings of the 10th International Cross-Disciplinary Conference on Web Accessibility}} (Rio de Janeiro, Brazil) \emph{(\bibinfo{series}{W4A '13})}. \bibinfo{publisher}{Association for Computing Machinery}, \bibinfo{address}{New York, NY, USA}, Article \bibinfo{articleno}{16}, \bibinfo{numpages}{4}~pages.
\newblock
\showISBNx{9781450318440}
\urldef\tempurl%
\url{https://doi.org/10.1145/2461121.2461137}
\showDOI{\tempurl}


\bibitem[Described and Program(2020)]%
        {DCMP}
\bibfield{author}{\bibinfo{person}{Described} {and} \bibinfo{person}{Captioned~Media Program}.} \bibinfo{year}{2020}\natexlab{}.
\newblock \bibinfo{title}{Described and Captioned Media Program (DCMP)}.
\newblock \bibinfo{howpublished}{\url{http://www.descriptionkey.org/quality_description.html}}.
\newblock
\newblock
\shownote{Accessed: 2019-03-19}.


\bibitem[Disney+(2023)]%
        {disney}
\bibfield{author}{\bibinfo{person}{Disney+}.} \bibinfo{year}{2023}\natexlab{}.
\newblock \bibinfo{title}{Disney+}.
\newblock \bibinfo{howpublished}{\url{http://www.disneyplus.com}}.
\newblock
\newblock
\shownote{Accessed: 2023-09-14}.


\bibitem[Doe(2016)]%
        {lionKing}
\bibfield{author}{\bibinfo{person}{Jane Doe}.} \bibinfo{year}{2016}\natexlab{}.
\newblock \bibinfo{title}{Audio Description - Full Clip}.
\newblock \bibinfo{howpublished}{\url{https://youtu.be/7-XOHN2BWG4}}.
\newblock
\newblock
\shownote{Accessed: 2023-11-6}.


\bibitem[Drisko and Maschi(2016)]%
        {drisko2016content}
\bibfield{author}{\bibinfo{person}{James~W Drisko} {and} \bibinfo{person}{Tina Maschi}.} \bibinfo{year}{2016}\natexlab{}.
\newblock \bibinfo{booktitle}{\emph{Content analysis}}.
\newblock \bibinfo{publisher}{Oxford University Press, USA}.
\newblock


\bibitem[Eyes(2024)]%
        {BeMyEyes}
\bibfield{author}{\bibinfo{person}{Be~My Eyes}.} \bibinfo{year}{2024}\natexlab{}.
\newblock \bibinfo{title}{Be My Eyes}.
\newblock \bibinfo{howpublished}{\url{https://www.bemyeyes.com/}}.
\newblock
\newblock
\shownote{Accessed: 2024-09-24}.


\bibitem[Food and Service(2022)]%
        {cheeseTostada}
\bibfield{author}{\bibinfo{person}{USDA Food} {and} \bibinfo{person}{Nutrition Service}.} \bibinfo{year}{2022}\natexlab{}.
\newblock \bibinfo{title}{CACFP Cooking Video: Cheesy Bean Tostada Ages 6-18, with Audio Description}.
\newblock \bibinfo{howpublished}{\url{https://youtu.be/9H8Ch1tcaCs}}.
\newblock
\newblock
\shownote{Accessed: 2023-11-6}.


\bibitem[for~the Blind(2024)]%
        {BeSpecular}
\bibfield{author}{\bibinfo{person}{American~Foundation for~the Blind}.} \bibinfo{year}{2024}\natexlab{}.
\newblock \bibinfo{title}{BeSpecular: A New Remote Assitant Service}.
\newblock \bibinfo{howpublished}{\url{https://www.afb.org/aw/17/7/15313}}.
\newblock
\newblock
\shownote{Accessed: 2024-09-24}.


\bibitem[Fryer(2016)]%
        {fryer2016introduction}
\bibfield{author}{\bibinfo{person}{Louise Fryer}.} \bibinfo{year}{2016}\natexlab{}.
\newblock \bibinfo{booktitle}{\emph{An introduction to audio description: A practical guide}}.
\newblock \bibinfo{publisher}{Routledge}.
\newblock


\bibitem[Gagnon et~al\mbox{.}(2010)]%
        {gagnon2010computer}
\bibfield{author}{\bibinfo{person}{L. Gagnon}, \bibinfo{person}{C. Chapdelaine}, \bibinfo{person}{D. Byrns}, \bibinfo{person}{S. Foucher}, \bibinfo{person}{M. Héritier}, {and} \bibinfo{person}{V. Gupta}.} \bibinfo{year}{2010}\natexlab{}.
\newblock \showarticletitle{A computer-vision-assisted system for Videodescription scripting}. In \bibinfo{booktitle}{\emph{2010 IEEE Computer Society Conference on Computer Vision and Pattern Recognition - Workshops}}. \bibinfo{pages}{41--48}.
\newblock
\urldef\tempurl%
\url{https://doi.org/10.1109/CVPRW.2010.5543575}
\showDOI{\tempurl}


\bibitem[Gajos and Weld(2004)]%
        {gajos2004supple}
\bibfield{author}{\bibinfo{person}{Krzysztof Gajos} {and} \bibinfo{person}{Daniel~S. Weld}.} \bibinfo{year}{2004}\natexlab{}.
\newblock \showarticletitle{SUPPLE: automatically generating user interfaces}. In \bibinfo{booktitle}{\emph{Proceedings of the 9th International Conference on Intelligent User Interfaces}} (Funchal, Madeira, Portugal) \emph{(\bibinfo{series}{IUI '04})}. \bibinfo{publisher}{Association for Computing Machinery}, \bibinfo{address}{New York, NY, USA}, \bibinfo{pages}{93–100}.
\newblock
\showISBNx{1581138156}
\urldef\tempurl%
\url{https://doi.org/10.1145/964442.964461}
\showDOI{\tempurl}


\bibitem[Global(2023)]%
        {sonyAds}
\bibfield{author}{\bibinfo{person}{Sony Global}.} \bibinfo{year}{2023}\natexlab{}.
\newblock \bibinfo{title}{Sony’s Purpose (with Audio Description) | Official Video}.
\newblock \bibinfo{howpublished}{\url{https://youtu.be/7Tiem2QBS0U}}.
\newblock
\newblock
\shownote{Accessed: 2023-11-6}.


\bibitem[Goodman(1961)]%
        {goodman1961snowball}
\bibfield{author}{\bibinfo{person}{Leo~A Goodman}.} \bibinfo{year}{1961}\natexlab{}.
\newblock \showarticletitle{Snowball sampling}.
\newblock \bibinfo{journal}{\emph{The annals of mathematical statistics}} (\bibinfo{year}{1961}), \bibinfo{pages}{148--170}.
\newblock


\bibitem[Google(2023)]%
        {googleInsight}
\bibfield{author}{\bibinfo{person}{Google}.} \bibinfo{year}{2023}\natexlab{}.
\newblock \bibinfo{title}{The Latest YouTube Stats on When, Where, and What people watch}.
\newblock \bibinfo{howpublished}{\url{https://www.thinkwithgoogle.com/consumer-insights/consumer-trends/youtube-stats-video-consumption-trends/}}.
\newblock
\newblock
\shownote{Accessed: 2023-09-12}.


\bibitem[Greening and Rolph(2007)]%
        {greening2007accessibility}
\bibfield{author}{\bibinfo{person}{Joan Greening} {and} \bibinfo{person}{Deborah Rolph}.} \bibinfo{year}{2007}\natexlab{}.
\newblock \showarticletitle{Accessibility: raising awareness of audio description in the UK}.
\newblock In \bibinfo{booktitle}{\emph{Media for All}}. \bibinfo{publisher}{Brill}, \bibinfo{pages}{127--138}.
\newblock


\bibitem[Gurari et~al\mbox{.}(2020)]%
        {gurari2020captioning}
\bibfield{author}{\bibinfo{person}{Danna Gurari}, \bibinfo{person}{Yinan Zhao}, \bibinfo{person}{Meng Zhang}, {and} \bibinfo{person}{Nilavra Bhattacharya}.} \bibinfo{year}{2020}\natexlab{}.
\newblock \showarticletitle{Captioning Images Taken by People Who Are Blind}. In \bibinfo{booktitle}{\emph{Computer Vision -- ECCV 2020}}, \bibfield{editor}{\bibinfo{person}{Andrea Vedaldi}, \bibinfo{person}{Horst Bischof}, \bibinfo{person}{Thomas Brox}, {and} \bibinfo{person}{Jan-Michael Frahm}} (Eds.). \bibinfo{publisher}{Springer International Publishing}, \bibinfo{address}{Cham}, \bibinfo{pages}{417--434}.
\newblock
\showISBNx{978-3-030-58520-4}


\bibitem[Halsted and Roberts(2002)]%
        {halsted2002eclipse}
\bibfield{author}{\bibinfo{person}{Kari Halsted} {and} \bibinfo{person}{James Roberts}.} \bibinfo{year}{2002}\natexlab{}.
\newblock \showarticletitle{Eclipse help system: an open source user assistance offering}. In \bibinfo{booktitle}{\emph{Proceedings of the 20th annual international conference on Computer documentation}}. \bibinfo{pages}{49--59}.
\newblock
\urldef\tempurl%
\url{https://doi.org/10.1145/584955.584964}
\showDOI{\tempurl}


\bibitem[Heap et~al\mbox{.}(2023)]%
        {heap2023experiencing}
\bibfield{author}{\bibinfo{person}{Tania Heap}, \bibinfo{person}{Regina Kaplan-Rakowski}, {and} \bibinfo{person}{Audon Archibald}.} \bibinfo{year}{2023}\natexlab{}.
\newblock \showarticletitle{Experiencing Virtual Reality for Perspective-Taking of Blind and Visually Impaired Learners}.
\newblock \bibinfo{journal}{\emph{Available at SSRN 4595370}} (\bibinfo{year}{2023}).
\newblock


\bibitem[Herskovitz et~al\mbox{.}(2023)]%
        {herskovitz2023hacking}
\bibfield{author}{\bibinfo{person}{Jaylin Herskovitz}, \bibinfo{person}{Andi Xu}, \bibinfo{person}{Rahaf Alharbi}, {and} \bibinfo{person}{Anhong Guo}.} \bibinfo{year}{2023}\natexlab{}.
\newblock \showarticletitle{Hacking, Switching, Combining: Understanding and Supporting DIY Assistive Technology Design by Blind People}. In \bibinfo{booktitle}{\emph{Proceedings of the 2023 CHI Conference on Human Factors in Computing Systems}} (Hamburg, Germany) \emph{(\bibinfo{series}{CHI '23})}. \bibinfo{publisher}{Association for Computing Machinery}, \bibinfo{address}{New York, NY, USA}, Article \bibinfo{articleno}{57}, \bibinfo{numpages}{17}~pages.
\newblock
\showISBNx{9781450394215}
\urldef\tempurl%
\url{https://doi.org/10.1145/3544548.3581249}
\showDOI{\tempurl}


\bibitem[Hruschka et~al\mbox{.}(2004)]%
        {hruschka2004reliability}
\bibfield{author}{\bibinfo{person}{Daniel~J Hruschka}, \bibinfo{person}{Deborah Schwartz}, \bibinfo{person}{Daphne~Cobb St.~John}, \bibinfo{person}{Erin Picone-Decaro}, \bibinfo{person}{Richard~A Jenkins}, {and} \bibinfo{person}{James~W Carey}.} \bibinfo{year}{2004}\natexlab{}.
\newblock \showarticletitle{Reliability in coding open-ended data: Lessons learned from HIV behavioral research}.
\newblock \bibinfo{journal}{\emph{Field methods}} \bibinfo{volume}{16}, \bibinfo{number}{3} (\bibinfo{year}{2004}), \bibinfo{pages}{307--331}.
\newblock
\urldef\tempurl%
\url{https://doi.org/10.1177/1525822X04266540}
\showDOI{\tempurl}


\bibitem[Huh et~al\mbox{.}(2023a)]%
        {huh2023genassist}
\bibfield{author}{\bibinfo{person}{Mina Huh}, \bibinfo{person}{Yi-Hao Peng}, {and} \bibinfo{person}{Amy Pavel}.} \bibinfo{year}{2023}\natexlab{a}.
\newblock \showarticletitle{GenAssist: Making Image Generation Accessible}. In \bibinfo{booktitle}{\emph{Proceedings of the 36th Annual ACM Symposium on User Interface Software and Technology}} (San Francisco, CA, USA) \emph{(\bibinfo{series}{UIST '23})}. \bibinfo{publisher}{Association for Computing Machinery}, \bibinfo{address}{New York, NY, USA}, Article \bibinfo{articleno}{38}, \bibinfo{numpages}{17}~pages.
\newblock
\showISBNx{9798400701320}
\urldef\tempurl%
\url{https://doi.org/10.1145/3586183.3606735}
\showDOI{\tempurl}


\bibitem[Huh et~al\mbox{.}(2023b)]%
        {huh2023avscript}
\bibfield{author}{\bibinfo{person}{Mina Huh}, \bibinfo{person}{Saelyne Yang}, \bibinfo{person}{Yi-Hao Peng}, \bibinfo{person}{Xiang~'Anthony' Chen}, \bibinfo{person}{Young-Ho Kim}, {and} \bibinfo{person}{Amy Pavel}.} \bibinfo{year}{2023}\natexlab{b}.
\newblock \showarticletitle{AVscript: Accessible Video Editing with Audio-Visual Scripts}. In \bibinfo{booktitle}{\emph{Proceedings of the 2023 CHI Conference on Human Factors in Computing Systems}} (Hamburg, Germany) \emph{(\bibinfo{series}{CHI '23})}. \bibinfo{publisher}{Association for Computing Machinery}, \bibinfo{address}{New York, NY, USA}, Article \bibinfo{articleno}{796}, \bibinfo{numpages}{17}~pages.
\newblock
\showISBNx{9781450394215}
\urldef\tempurl%
\url{https://doi.org/10.1145/3544548.3581494}
\showDOI{\tempurl}


\bibitem[Hurst et~al\mbox{.}(2007)]%
        {hurst2007dynamic}
\bibfield{author}{\bibinfo{person}{Amy Hurst}, \bibinfo{person}{Scott~E. Hudson}, {and} \bibinfo{person}{Jennifer Mankoff}.} \bibinfo{year}{2007}\natexlab{}.
\newblock \showarticletitle{Dynamic detection of novice vs. skilled use without a task model}. In \bibinfo{booktitle}{\emph{Proceedings of the SIGCHI Conference on Human Factors in Computing Systems}} (San Jose, California, USA) \emph{(\bibinfo{series}{CHI '07})}. \bibinfo{publisher}{Association for Computing Machinery}, \bibinfo{address}{New York, NY, USA}, \bibinfo{pages}{271–280}.
\newblock
\showISBNx{9781595935939}
\urldef\tempurl%
\url{https://doi.org/10.1145/1240624.1240669}
\showDOI{\tempurl}


\bibitem[IMSTVUK(2013)]%
        {en2}
\bibfield{author}{\bibinfo{person}{IMSTVUK}.} \bibinfo{year}{2013}\natexlab{}.
\newblock \bibinfo{title}{Frozen - Trailer with Audio Description)}.
\newblock \bibinfo{howpublished}{\url{https://youtu.be/O7j4_aP8dWA}}.
\newblock
\newblock
\shownote{Accessed: 2023-11-6}.


\bibitem[Initiative(2023)]%
        {w3c-wai}
\bibfield{author}{\bibinfo{person}{World-Wide Web COnsortium Web~Accessibility Initiative}.} \bibinfo{year}{2023}\natexlab{}.
\newblock \bibinfo{title}{Making the Web-Accessible}.
\newblock \bibinfo{howpublished}{\url{https://www.w3.org/WAI/}}.
\newblock
\newblock
\shownote{Accessed: 2023-11-6}.


\bibitem[Jiang et~al\mbox{.}(2024)]%
        {jiang2024it}
\bibfield{author}{\bibinfo{person}{Lucy Jiang}, \bibinfo{person}{Crescentia Jung}, \bibinfo{person}{Mahika Phutane}, \bibinfo{person}{Abigale Stangl}, {and} \bibinfo{person}{Shiri Azenkot}.} \bibinfo{year}{2024}\natexlab{}.
\newblock \showarticletitle{“It’s Kind of Context Dependent”: Understanding Blind and Low Vision People’s Video Accessibility Preferences Across Viewing Scenarios}. In \bibinfo{booktitle}{\emph{Proceedings of the CHI Conference on Human Factors in Computing Systems}} (Honolulu, HI, USA) \emph{(\bibinfo{series}{CHI '24})}. \bibinfo{publisher}{Association for Computing Machinery}, \bibinfo{address}{New York, NY, USA}, Article \bibinfo{articleno}{897}, \bibinfo{numpages}{20}~pages.
\newblock
\showISBNx{9798400703300}
\urldef\tempurl%
\url{https://doi.org/10.1145/3613904.3642238}
\showDOI{\tempurl}


\bibitem[Jiang and Ladner(2022)]%
        {jiang2022co}
\bibfield{author}{\bibinfo{person}{Lucy Jiang} {and} \bibinfo{person}{Richard Ladner}.} \bibinfo{year}{2022}\natexlab{}.
\newblock \showarticletitle{Co-Designing Systems to Support Blind and Low Vision Audio Description Writers}. In \bibinfo{booktitle}{\emph{Proceedings of the 24th International ACM SIGACCESS Conference on Computers and Accessibility}} (Athens, Greece) \emph{(\bibinfo{series}{ASSETS '22})}. \bibinfo{publisher}{Association for Computing Machinery}, \bibinfo{address}{New York, NY, USA}, Article \bibinfo{articleno}{74}, \bibinfo{numpages}{3}~pages.
\newblock
\showISBNx{9781450392587}
\urldef\tempurl%
\url{https://doi.org/10.1145/3517428.3550394}
\showDOI{\tempurl}


\bibitem[Jiang et~al\mbox{.}(2023)]%
        {jiang2023beyond}
\bibfield{author}{\bibinfo{person}{Lucy Jiang}, \bibinfo{person}{Mahika Phutane}, {and} \bibinfo{person}{Shiri Azenkot}.} \bibinfo{year}{2023}\natexlab{}.
\newblock \showarticletitle{Beyond Audio Description: Exploring 360° Video Accessibility with Blind and Low Vision Users Through Collaborative Creation}. In \bibinfo{booktitle}{\emph{Proceedings of the 25th International ACM SIGACCESS Conference on Computers and Accessibility}} (New York, NY, USA) \emph{(\bibinfo{series}{ASSETS '23})}. \bibinfo{publisher}{Association for Computing Machinery}, \bibinfo{address}{New York, NY, USA}, Article \bibinfo{articleno}{50}, \bibinfo{numpages}{17}~pages.
\newblock
\showISBNx{9798400702204}
\urldef\tempurl%
\url{https://doi.org/10.1145/3597638.3608381}
\showDOI{\tempurl}


\bibitem[Jordan et~al\mbox{.}(1996)]%
        {jordan1996usability}
\bibfield{author}{\bibinfo{person}{Patrick~W Jordan}, \bibinfo{person}{Bruce Thomas}, \bibinfo{person}{Ian~Lyall McClelland}, {and} \bibinfo{person}{Bernard Weerdmeester}.} \bibinfo{year}{1996}\natexlab{}.
\newblock \bibinfo{booktitle}{\emph{Usability evaluation in industry}}.
\newblock \bibinfo{publisher}{CRC Press}.
\newblock


\bibitem[Katsarova(2018)]%
        {katsarova2018audiovisual}
\bibfield{author}{\bibinfo{person}{Ivana Katsarova}.} \bibinfo{year}{2018}\natexlab{}.
\newblock \showarticletitle{The audiovisual media services directive}.
\newblock \bibinfo{journal}{\emph{Briefing EU [Legislation in Progress,] European Parliament}} (\bibinfo{year}{2018}).
\newblock


\bibitem[Kegishyan(2023)]%
        {yansmedia}
\bibfield{author}{\bibinfo{person}{Irina Kegishyan}.} \bibinfo{year}{2023}\natexlab{}.
\newblock \bibinfo{title}{Mobile Video Statistics}.
\newblock \bibinfo{howpublished}{\url{https://www.yansmedia.com/blog/mobile-video-statistics}}.
\newblock
\newblock
\shownote{Accessed: 2023-04-09}.


\bibitem[Kennedy-Eden and Gretzel(2012)]%
        {kennedy2012taxonomy}
\bibfield{author}{\bibinfo{person}{Heather Kennedy-Eden} {and} \bibinfo{person}{Ulrike Gretzel}.} \bibinfo{year}{2012}\natexlab{}.
\newblock \showarticletitle{A Taxonomy of Mobile Applications in Tourism}.
\newblock \bibinfo{journal}{\emph{e-Review of Tourism Research (eRTR)}}  \bibinfo{volume}{10} (\bibinfo{date}{01} \bibinfo{year}{2012}), \bibinfo{pages}{47--50}.
\newblock


\bibitem[Kitchen(2021)]%
        {tut1}
\bibfield{author}{\bibinfo{person}{Em's Kitchen}.} \bibinfo{year}{2021}\natexlab{}.
\newblock \bibinfo{title}{3 Ingredient Nutella Mug Cake 2 Ways}.
\newblock \bibinfo{howpublished}{\url{https://youtu.be/sItYaC1z_d0}}.
\newblock
\newblock
\shownote{Accessed: 2023-11-6}.


\bibitem[Kobayashi et~al\mbox{.}(2009)]%
        {kobayashi2009providing}
\bibfield{author}{\bibinfo{person}{Masatomo Kobayashi}, \bibinfo{person}{Kentarou Fukuda}, \bibinfo{person}{Hironobu Takagi}, {and} \bibinfo{person}{Chieko Asakawa}.} \bibinfo{year}{2009}\natexlab{}.
\newblock \showarticletitle{Providing synthesized audio description for online videos}. In \bibinfo{booktitle}{\emph{Proceedings of the 11th International ACM SIGACCESS Conference on Computers and Accessibility}} (Pittsburgh, Pennsylvania, USA) \emph{(\bibinfo{series}{ASSETS '09})}. \bibinfo{publisher}{Association for Computing Machinery}, \bibinfo{address}{New York, NY, USA}, \bibinfo{pages}{249–250}.
\newblock
\showISBNx{9781605585581}
\urldef\tempurl%
\url{https://doi.org/10.1145/1639642.1639699}
\showDOI{\tempurl}


\bibitem[Lei et~al\mbox{.}(2020)]%
        {lei2020mart}
\bibfield{author}{\bibinfo{person}{Jie Lei}, \bibinfo{person}{Liwei Wang}, \bibinfo{person}{Yelong Shen}, \bibinfo{person}{Dong Yu}, \bibinfo{person}{Tamara Berg}, {and} \bibinfo{person}{Mohit Bansal}.} \bibinfo{year}{2020}\natexlab{}.
\newblock \showarticletitle{{MART}: Memory-Augmented Recurrent Transformer for Coherent Video Paragraph Captioning}. In \bibinfo{booktitle}{\emph{Proceedings of the 58th Annual Meeting of the Association for Computational Linguistics}}, \bibfield{editor}{\bibinfo{person}{Dan Jurafsky}, \bibinfo{person}{Joyce Chai}, \bibinfo{person}{Natalie Schluter}, {and} \bibinfo{person}{Joel Tetreault}} (Eds.). \bibinfo{publisher}{Association for Computational Linguistics}, \bibinfo{address}{Online}, \bibinfo{pages}{2603--2614}.
\newblock
\urldef\tempurl%
\url{https://doi.org/10.18653/v1/2020.acl-main.233}
\showDOI{\tempurl}


\bibitem[Leung(2018)]%
        {leung2018audio}
\bibfield{author}{\bibinfo{person}{Hoi Ching~Dawning Leung}.} \bibinfo{year}{2018}\natexlab{}.
\newblock \emph{\bibinfo{title}{Audio description of audiovisual programmes for the visually impaired in Hong Kong}}.
\newblock \bibinfo{thesistype}{Ph.\,D. Dissertation}. \bibinfo{school}{UCL (University College London)}.
\newblock


\bibitem[Liu et~al\mbox{.}(2021)]%
        {liu2021makes}
\bibfield{author}{\bibinfo{person}{Xingyu Liu}, \bibinfo{person}{Patrick Carrington}, \bibinfo{person}{Xiang~'Anthony' Chen}, {and} \bibinfo{person}{Amy Pavel}.} \bibinfo{year}{2021}\natexlab{}.
\newblock \showarticletitle{What Makes Videos Accessible to Blind and Visually Impaired People?}. In \bibinfo{booktitle}{\emph{Proceedings of the 2021 CHI Conference on Human Factors in Computing Systems}} (Yokohama, Japan) \emph{(\bibinfo{series}{CHI '21})}. \bibinfo{publisher}{Association for Computing Machinery}, \bibinfo{address}{New York, NY, USA}, Article \bibinfo{articleno}{272}, \bibinfo{numpages}{14}~pages.
\newblock
\showISBNx{9781450380966}
\urldef\tempurl%
\url{https://doi.org/10.1145/3411764.3445233}
\showDOI{\tempurl}


\bibitem[Lo~Valvo et~al\mbox{.}(2021)]%
        {lo2021navigation}
\bibfield{author}{\bibinfo{person}{Alice Lo~Valvo}, \bibinfo{person}{Daniele Croce}, \bibinfo{person}{Domenico Garlisi}, \bibinfo{person}{Fabrizio Giuliano}, \bibinfo{person}{Laura Giarr{\'e}}, {and} \bibinfo{person}{Ilenia Tinnirello}.} \bibinfo{year}{2021}\natexlab{}.
\newblock \showarticletitle{A navigation and augmented reality system for visually impaired people}.
\newblock \bibinfo{journal}{\emph{Sensors}} \bibinfo{volume}{21}, \bibinfo{number}{9} (\bibinfo{year}{2021}), \bibinfo{pages}{3061}.
\newblock
\urldef\tempurl%
\url{https://doi.org/10.3390/s21093061}
\showDOI{\tempurl}


\bibitem[Lopez et~al\mbox{.}(2018)]%
        {lopez2018audio}
\bibfield{author}{\bibinfo{person}{Mariana Lopez}, \bibinfo{person}{Gavin Kearney}, {and} \bibinfo{person}{Kriszti{\'a}n Hofst{\"a}dter}.} \bibinfo{year}{2018}\natexlab{}.
\newblock \showarticletitle{Audio Description in the UK: What works, what doesn’t, and understanding the need for personalising access}.
\newblock \bibinfo{journal}{\emph{British journal of visual impairment}} \bibinfo{volume}{36}, \bibinfo{number}{3} (\bibinfo{year}{2018}), \bibinfo{pages}{274--291}.
\newblock


\bibitem[Microsoft(2024)]%
        {Soundscape}
\bibfield{author}{\bibinfo{person}{Microsoft}.} \bibinfo{year}{2024}\natexlab{}.
\newblock \bibinfo{title}{Miscrosoft Soundscape: A map delivered in 3D Sound}.
\newblock \bibinfo{howpublished}{\url{https://www.microsoft.com/en-us/research/product/soundscape/}}.
\newblock
\newblock
\shownote{Accessed: 2024-09-24}.


\bibitem[Montagud et~al\mbox{.}(2020)]%
        {montagud2020culture}
\bibfield{author}{\bibinfo{person}{Mario Montagud}, \bibinfo{person}{Pilar Orero}, {and} \bibinfo{person}{Anna Matamala}.} \bibinfo{year}{2020}\natexlab{}.
\newblock \showarticletitle{Culture 4 all: accessibility-enabled cultural experiences through immersive VR360 content}.
\newblock \bibinfo{journal}{\emph{Personal and Ubiquitous Computing}} \bibinfo{volume}{24}, \bibinfo{number}{6} (\bibinfo{year}{2020}), \bibinfo{pages}{887--905}.
\newblock
\urldef\tempurl%
\url{https://doi.org/10.1007/s00779-019-01357-3}
\showDOI{\tempurl}


\bibitem[Morley(1998)]%
        {morley1998digital}
\bibfield{author}{\bibinfo{person}{Sarah Morley}.} \bibinfo{year}{1998}\natexlab{}.
\newblock \showarticletitle{Digital talking books on a PC: a usability evaluation of the prototype DAISY playback software}. In \bibinfo{booktitle}{\emph{Proceedings of the Third International ACM Conference on Assistive Technologies}} (Marina del Rey, California, USA) \emph{(\bibinfo{series}{Assets '98})}. \bibinfo{publisher}{Association for Computing Machinery}, \bibinfo{address}{New York, NY, USA}, \bibinfo{pages}{157–164}.
\newblock
\showISBNx{1581130201}
\urldef\tempurl%
\url{https://doi.org/10.1145/274497.274527}
\showDOI{\tempurl}


\bibitem[Natalie et~al\mbox{.}(2020)]%
        {natalie2020viscene}
\bibfield{author}{\bibinfo{person}{Rosiana Natalie}, \bibinfo{person}{Ebrima Jarjue}, \bibinfo{person}{Hernisa Kacorri}, {and} \bibinfo{person}{Kotaro Hara}.} \bibinfo{year}{2020}\natexlab{}.
\newblock \showarticletitle{ViScene: A Collaborative Authoring Tool for Scene Descriptions in Videos}. In \bibinfo{booktitle}{\emph{Proceedings of the 22nd International ACM SIGACCESS Conference on Computers and Accessibility}} (Virtual Event, Greece) \emph{(\bibinfo{series}{ASSETS '20})}. \bibinfo{publisher}{Association for Computing Machinery}, \bibinfo{address}{New York, NY, USA}, Article \bibinfo{articleno}{87}, \bibinfo{numpages}{4}~pages.
\newblock
\showISBNx{9781450371032}
\urldef\tempurl%
\url{https://doi.org/10.1145/3373625.3418030}
\showDOI{\tempurl}


\bibitem[Natalie et~al\mbox{.}(2021)]%
        {natalie2021efficacy}
\bibfield{author}{\bibinfo{person}{Rosiana Natalie}, \bibinfo{person}{Jolene Loh}, \bibinfo{person}{Huei~Suen Tan}, \bibinfo{person}{Joshua Tseng}, \bibinfo{person}{Ian Luke Yi-Ren Chan}, \bibinfo{person}{Ebrima~H Jarjue}, \bibinfo{person}{Hernisa Kacorri}, {and} \bibinfo{person}{Kotaro Hara}.} \bibinfo{year}{2021}\natexlab{}.
\newblock \showarticletitle{The Efficacy of Collaborative Authoring of Video Scene Descriptions}. In \bibinfo{booktitle}{\emph{Proceedings of the 23rd International ACM SIGACCESS Conference on Computers and Accessibility}} (Virtual Event, USA) \emph{(\bibinfo{series}{ASSETS '21})}. \bibinfo{publisher}{Association for Computing Machinery}, \bibinfo{address}{New York, NY, USA}, Article \bibinfo{articleno}{17}, \bibinfo{numpages}{15}~pages.
\newblock
\showISBNx{9781450383066}
\urldef\tempurl%
\url{https://doi.org/10.1145/3441852.3471201}
\showDOI{\tempurl}


\bibitem[Natalie et~al\mbox{.}(2023)]%
        {natalie2023supporting}
\bibfield{author}{\bibinfo{person}{Rosiana Natalie}, \bibinfo{person}{Joshua Tseng}, \bibinfo{person}{Hernisa Kacorri}, {and} \bibinfo{person}{Kotaro Hara}.} \bibinfo{year}{2023}\natexlab{}.
\newblock \showarticletitle{Supporting Novices Author Audio Descriptions via Automatic Feedback}. In \bibinfo{booktitle}{\emph{Proceedings of the 2023 CHI Conference on Human Factors in Computing Systems}} (Hamburg, Germany) \emph{(\bibinfo{series}{CHI '23})}. \bibinfo{publisher}{Association for Computing Machinery}, \bibinfo{address}{New York, NY, USA}, Article \bibinfo{articleno}{77}, \bibinfo{numpages}{18}~pages.
\newblock
\showISBNx{9781450394215}
\urldef\tempurl%
\url{https://doi.org/10.1145/3544548.3581023}
\showDOI{\tempurl}


\bibitem[Netflix(2020)]%
        {netflixADGuide}
\bibfield{author}{\bibinfo{person}{Netflix}.} \bibinfo{year}{2020}\natexlab{}.
\newblock \bibinfo{title}{Audio Description Style Guide v2.1}.
\newblock \bibinfo{howpublished}{\url{https://partnerhelp.netflixstudios.com/hc/en-us/articles/215510667-Audio-Description-Style-Guide-v2-1}}.
\newblock
\newblock
\shownote{Accessed: 2020-11-6}.


\bibitem[Netflix(2023)]%
        {netflix}
\bibfield{author}{\bibinfo{person}{Netflix}.} \bibinfo{year}{2023}\natexlab{}.
\newblock \bibinfo{title}{Netflix}.
\newblock \bibinfo{howpublished}{\url{http://www.netflix.com}}.
\newblock
\newblock
\shownote{Accessed: 2023-09-14}.


\bibitem[Ni et~al\mbox{.}(2022)]%
        {ni2022expanding}
\bibfield{author}{\bibinfo{person}{Bolin Ni}, \bibinfo{person}{Houwen Peng}, \bibinfo{person}{Minghao Chen}, \bibinfo{person}{Songyang Zhang}, \bibinfo{person}{Gaofeng Meng}, \bibinfo{person}{Jianlong Fu}, \bibinfo{person}{Shiming Xiang}, {and} \bibinfo{person}{Haibin Ling}.} \bibinfo{year}{2022}\natexlab{}.
\newblock \showarticletitle{Expanding Language-Image Pretrained Models for General Video Recognition}. In \bibinfo{booktitle}{\emph{Computer Vision -- ECCV 2022}}, \bibfield{editor}{\bibinfo{person}{Shai Avidan}, \bibinfo{person}{Gabriel Brostow}, \bibinfo{person}{Moustapha Ciss{\'e}}, \bibinfo{person}{Giovanni~Maria Farinella}, {and} \bibinfo{person}{Tal Hassner}} (Eds.). \bibinfo{publisher}{Springer Nature Switzerland}, \bibinfo{address}{Cham}, \bibinfo{pages}{1--18}.
\newblock
\showISBNx{978-3-031-19772-7}


\bibitem[Pavel et~al\mbox{.}(2020)]%
        {pavel2020rescribe}
\bibfield{author}{\bibinfo{person}{Amy Pavel}, \bibinfo{person}{Gabriel Reyes}, {and} \bibinfo{person}{Jeffrey~P. Bigham}.} \bibinfo{year}{2020}\natexlab{}.
\newblock \showarticletitle{Rescribe: Authoring and Automatically Editing Audio Descriptions}. In \bibinfo{booktitle}{\emph{Proceedings of the 33rd Annual ACM Symposium on User Interface Software and Technology}} (Virtual Event, USA) \emph{(\bibinfo{series}{UIST '20})}. \bibinfo{publisher}{Association for Computing Machinery}, \bibinfo{address}{New York, NY, USA}, \bibinfo{pages}{747–759}.
\newblock
\showISBNx{9781450375146}
\urldef\tempurl%
\url{https://doi.org/10.1145/3379337.3415864}
\showDOI{\tempurl}


\bibitem[Pawlowski(2010)]%
        {pawlowski2010basic}
\bibfield{author}{\bibinfo{person}{Peter Pawlowski}.} \bibinfo{year}{2010}\natexlab{}.
\newblock \showarticletitle{Basic Player Whose Appearance and Functions can be Customized Freely ‘Foobar 2000’v1. 0 is Unveiled}.
\newblock \bibinfo{journal}{\emph{Windows Forest, Japan, Jan}}  \bibinfo{volume}{12} (\bibinfo{year}{2010}), \bibinfo{pages}{3}.
\newblock


\bibitem[Pictory(2023)]%
        {pictory}
\bibfield{author}{\bibinfo{person}{Pictory}.} \bibinfo{year}{2023}\natexlab{}.
\newblock \bibinfo{title}{What are the Most Popular Genres on YouTube in 2023?}
\newblock \bibinfo{howpublished}{\url{https://pictory.ai/blog/what-are-the-most-popular-genres-on-youtube-in-2023?el=0035&htrafficsource=pictoryblog&hcategory=video}}.
\newblock
\newblock
\shownote{Accessed: 2024-04-18}.


\bibitem[Player(2020)]%
        {ableplayer}
\bibfield{author}{\bibinfo{person}{Able Player}.} \bibinfo{year}{2020}\natexlab{}.
\newblock \bibinfo{title}{Able Player: Fuly Accessible cross-browser HTML Media Player}.
\newblock \bibinfo{howpublished}{\url{https://www.3playmedia.com/services/features/plugins/3play-plugin/}}.
\newblock
\newblock
\shownote{Accessed: 2020-11-6}.


\bibitem[Ponsard and McGrenere(2016)]%
        {ponsard2016anchored}
\bibfield{author}{\bibinfo{person}{Antoine Ponsard} {and} \bibinfo{person}{Joanna McGrenere}.} \bibinfo{year}{2016}\natexlab{}.
\newblock \showarticletitle{Anchored Customization: Anchoring Settings to the Application Interface to Afford Customization}. In \bibinfo{booktitle}{\emph{Proceedings of the 2016 CHI Conference on Human Factors in Computing Systems}} (San Jose, California, USA) \emph{(\bibinfo{series}{CHI '16})}. \bibinfo{publisher}{Association for Computing Machinery}, \bibinfo{address}{New York, NY, USA}, \bibinfo{pages}{4154–4165}.
\newblock
\showISBNx{9781450333627}
\urldef\tempurl%
\url{https://doi.org/10.1145/2858036.2858129}
\showDOI{\tempurl}


\bibitem[Project(2023)]%
        {adp}
\bibfield{author}{\bibinfo{person}{Audio~Description Project}.} \bibinfo{year}{2023}\natexlab{}.
\newblock \bibinfo{title}{Recommendation of the Federal Communications Commission disability ...}
\newblock
\newblock
\urldef\tempurl%
\url{https://adp.acb.org/docs/DAC%20Recommendation%20on%20Audo%20Description%20Quality%20Adopted%20October%2014%202020.pdf}
\showURL{%
\tempurl}


\bibitem[Radford et~al\mbox{.}(2021)]%
        {radford2021learning}
\bibfield{author}{\bibinfo{person}{Alec Radford}, \bibinfo{person}{Jong~Wook Kim}, \bibinfo{person}{Chris Hallacy}, \bibinfo{person}{Aditya Ramesh}, \bibinfo{person}{Gabriel Goh}, \bibinfo{person}{Sandhini Agarwal}, \bibinfo{person}{Girish Sastry}, \bibinfo{person}{Amanda Askell}, \bibinfo{person}{Pamela Mishkin}, \bibinfo{person}{Jack Clark}, {et~al\mbox{.}}} \bibinfo{year}{2021}\natexlab{}.
\newblock \showarticletitle{Learning transferable visual models from natural language supervision}. In \bibinfo{booktitle}{\emph{International conference on machine learning}}. PMLR, \bibinfo{pages}{8748--8763}.
\newblock


\bibitem[Radford et~al\mbox{.}(2023)]%
        {radford2023robust}
\bibfield{author}{\bibinfo{person}{Alec Radford}, \bibinfo{person}{Jong~Wook Kim}, \bibinfo{person}{Tao Xu}, \bibinfo{person}{Greg Brockman}, \bibinfo{person}{Christine McLeavey}, {and} \bibinfo{person}{Ilya Sutskever}.} \bibinfo{year}{2023}\natexlab{}.
\newblock \showarticletitle{Robust speech recognition via large-scale weak supervision}. In \bibinfo{booktitle}{\emph{International Conference on Machine Learning}}. PMLR, \bibinfo{pages}{28492--28518}.
\newblock


\bibitem[Saati et~al\mbox{.}(2005)]%
        {saati2005towards}
\bibfield{author}{\bibinfo{person}{Batul Saati}, \bibinfo{person}{May Salem}, {and} \bibinfo{person}{Willem-Paul Brinkman}.} \bibinfo{year}{2005}\natexlab{}.
\newblock \showarticletitle{Towards customized user interface skins: investigating user personality and skin colour}.
\newblock \bibinfo{journal}{\emph{Proceedings of HCI 2005}}  \bibinfo{volume}{2} (\bibinfo{year}{2005}), \bibinfo{pages}{89--93}.
\newblock


\bibitem[Scientific(2023)]%
        {JAWS}
\bibfield{author}{\bibinfo{person}{Freedom Scientific}.} \bibinfo{year}{2023}\natexlab{}.
\newblock \bibinfo{title}{JAWS}.
\newblock \bibinfo{howpublished}{\url{https://www.freedomscientific.com/products/software/jaws/}}.
\newblock
\newblock
\shownote{Accessed: 2023-09-14}.


\bibitem[Stangl et~al\mbox{.}(2020)]%
        {stangl2020person}
\bibfield{author}{\bibinfo{person}{Abigale Stangl}, \bibinfo{person}{Meredith~Ringel Morris}, {and} \bibinfo{person}{Danna Gurari}.} \bibinfo{year}{2020}\natexlab{}.
\newblock \showarticletitle{"Person, Shoes, Tree. Is the Person Naked?" What People with Vision Impairments Want in Image Descriptions}. In \bibinfo{booktitle}{\emph{Proceedings of the 2020 CHI Conference on Human Factors in Computing Systems}} (Honolulu, HI, USA) \emph{(\bibinfo{series}{CHI '20})}. \bibinfo{publisher}{Association for Computing Machinery}, \bibinfo{address}{New York, NY, USA}, \bibinfo{pages}{1–13}.
\newblock
\showISBNx{9781450367080}
\urldef\tempurl%
\url{https://doi.org/10.1145/3313831.3376404}
\showDOI{\tempurl}


\bibitem[Stangl et~al\mbox{.}(2021)]%
        {stangl2021going}
\bibfield{author}{\bibinfo{person}{Abigale Stangl}, \bibinfo{person}{Nitin Verma}, \bibinfo{person}{Kenneth~R. Fleischmann}, \bibinfo{person}{Meredith~Ringel Morris}, {and} \bibinfo{person}{Danna Gurari}.} \bibinfo{year}{2021}\natexlab{}.
\newblock \showarticletitle{Going Beyond One-Size-Fits-All Image Descriptions to Satisfy the Information Wants of People Who are Blind or Have Low Vision}. In \bibinfo{booktitle}{\emph{Proceedings of the 23rd International ACM SIGACCESS Conference on Computers and Accessibility}} (Virtual Event, USA) \emph{(\bibinfo{series}{ASSETS '21})}. \bibinfo{publisher}{Association for Computing Machinery}, \bibinfo{address}{New York, NY, USA}, Article \bibinfo{articleno}{16}, \bibinfo{numpages}{15}~pages.
\newblock
\showISBNx{9781450383066}
\urldef\tempurl%
\url{https://doi.org/10.1145/3441852.3471233}
\showDOI{\tempurl}


\bibitem[Strober(2020)]%
        {imagineThat}
\bibfield{author}{\bibinfo{person}{Rena Strober}.} \bibinfo{year}{2020}\natexlab{}.
\newblock \bibinfo{title}{Imagine That! Music video with AUDIO DESCRIPTION}.
\newblock \bibinfo{howpublished}{\url{https://youtu.be/UXz9AtO_kl0}}.
\newblock
\newblock
\shownote{Accessed: 2023-11-6}.


\bibitem[Studios(2013)]%
        {en1}
\bibfield{author}{\bibinfo{person}{Walt Disney~Animation Studios}.} \bibinfo{year}{2013}\natexlab{}.
\newblock \bibinfo{title}{Disney's Frozen "Party Is Over" Clip)}.
\newblock \bibinfo{howpublished}{\url{https://youtu.be/jNuZC5_9pQQ}}.
\newblock
\newblock
\shownote{Accessed: 2023-11-6}.


\bibitem[Sundar and Marathe(2010)]%
        {sundar2010personalization}
\bibfield{author}{\bibinfo{person}{S~Shyam Sundar} {and} \bibinfo{person}{Sampada~S Marathe}.} \bibinfo{year}{2010}\natexlab{}.
\newblock \showarticletitle{Personalization versus customization: The importance of agency, privacy, and power usage}.
\newblock \bibinfo{journal}{\emph{Human communication research}} \bibinfo{volume}{36}, \bibinfo{number}{3} (\bibinfo{year}{2010}), \bibinfo{pages}{298--322}.
\newblock
\urldef\tempurl%
\url{https://doi.org/10.1111/j.1468-2958.2010.01377.x}
\showDOI{\tempurl}


\bibitem[Sung et~al\mbox{.}(2010)]%
        {sung2010social}
\bibfield{author}{\bibinfo{person}{Jieun Sung}, \bibinfo{person}{Torger Bjornrud}, \bibinfo{person}{Yu-hao Lee}, {and} \bibinfo{person}{D.~Yvette Wohn}.} \bibinfo{year}{2010}\natexlab{}.
\newblock \showarticletitle{Social network games: exploring audience traits}. In \bibinfo{booktitle}{\emph{CHI '10 Extended Abstracts on Human Factors in Computing Systems}} (Atlanta, Georgia, USA) \emph{(\bibinfo{series}{CHI EA '10})}. \bibinfo{publisher}{Association for Computing Machinery}, \bibinfo{address}{New York, NY, USA}, \bibinfo{pages}{3649–3654}.
\newblock
\showISBNx{9781605589305}
\urldef\tempurl%
\url{https://doi.org/10.1145/1753846.1754033}
\showDOI{\tempurl}


\bibitem[Sunkara et~al\mbox{.}(2023)]%
        {sunkara2023enabling}
\bibfield{author}{\bibinfo{person}{Mohan Sunkara}, \bibinfo{person}{Yash Prakash}, \bibinfo{person}{Hae-Na Lee}, \bibinfo{person}{Sampath Jayarathna}, {and} \bibinfo{person}{Vikas Ashok}.} \bibinfo{year}{2023}\natexlab{}.
\newblock \showarticletitle{Enabling Customization of Discussion Forums for Blind Users}.
\newblock \bibinfo{journal}{\emph{Proc. ACM Hum.-Comput. Interact.}} \bibinfo{volume}{7}, \bibinfo{number}{EICS}, Article \bibinfo{articleno}{176} (\bibinfo{date}{jun} \bibinfo{year}{2023}), \bibinfo{numpages}{20}~pages.
\newblock
\urldef\tempurl%
\url{https://doi.org/10.1145/3593228}
\showDOI{\tempurl}


\bibitem[Thompson(2019)]%
        {thompson2019}
\bibfield{author}{\bibinfo{person}{Terril Thompson}.} \bibinfo{year}{2019}\natexlab{}.
\newblock \bibinfo{title}{Audio Description using the Web Speech API}.
\newblock \bibinfo{howpublished}{\url{https://terrillthompson.com/1173}}.
\newblock
\newblock
\shownote{Accessed: 2020-11-6}.


\bibitem[Vera(2006)]%
        {vera2006translating}
\bibfield{author}{\bibinfo{person}{JF Vera}.} \bibinfo{year}{2006}\natexlab{}.
\newblock \showarticletitle{Translating audio description scripts: the way forward? Tentative first stage project results}. In \bibinfo{booktitle}{\emph{MuTra 2006 Audio Visual Translation Scenarios: Conference proceedings}}. \bibinfo{pages}{148--181}.
\newblock


\bibitem[W3C(2022)]%
        {w3c}
\bibfield{author}{\bibinfo{person}{W3C}.} \bibinfo{year}{2022}\natexlab{}.
\newblock \bibinfo{title}{Descriptions of Visual Information.}
\newblock \bibinfo{howpublished}{\url{https://www.w3.org/WAI/media/av/description/}}.
\newblock
\newblock
\shownote{Accessed: 2022-11-6}.


\bibitem[W3C(2023)]%
        {wcag}
\bibfield{author}{\bibinfo{person}{W3C}.} \bibinfo{year}{2023}\natexlab{}.
\newblock \bibinfo{title}{Extended Audio Description (Prerecorded) (Level AAA)}.
\newblock \bibinfo{howpublished}{\url{https://www.w3.org/TR/WCAG20-TECHS/G8.html. }}.
\newblock
\newblock
\shownote{Accessed: 2023-04-09}.


\bibitem[W3C(2024)]%
        {WCAG_SC12-7}
\bibfield{author}{\bibinfo{person}{W3C}.} \bibinfo{year}{2024}\natexlab{}.
\newblock \bibinfo{title}{Extended Audio Description (Prerecorded): Understanding SC 1.2.7}.
\newblock \bibinfo{howpublished}{\url{https://www.w3.org/TR/UNDERSTANDING-WCAG20/media-equiv-extended-ad.html}}.
\newblock
\newblock
\shownote{Accessed: 2024-03-14}.


\bibitem[(WAI)(2016a)]%
        {ex2}
\bibfield{author}{\bibinfo{person}{W3C Web Accessibility~Initiative (WAI)}.} \bibinfo{year}{2016}\natexlab{a}.
\newblock \bibinfo{title}{Web Accessibility Perspectives: Customizable Text - Audio Described Version}.
\newblock \bibinfo{howpublished}{\url{https://youtu.be/L4WLeVc5l5k}}.
\newblock
\newblock
\shownote{Accessed: 2023-11-6}.


\bibitem[(WAI)(2016b)]%
        {ex1}
\bibfield{author}{\bibinfo{person}{W3C Web Accessibility~Initiative (WAI)}.} \bibinfo{year}{2016}\natexlab{b}.
\newblock \bibinfo{title}{Web Accessibility Perspectives: Video Captions - Audio Described Version}.
\newblock \bibinfo{howpublished}{\url{https://youtu.be/4qIordU8vT8}}.
\newblock
\newblock
\shownote{Accessed: 2023-11-6}.


\bibitem[Walczak and Fryer(2018)]%
        {walczak2018vocal}
\bibfield{author}{\bibinfo{person}{Agnieszka Walczak} {and} \bibinfo{person}{Louise Fryer}.} \bibinfo{year}{2018}\natexlab{}.
\newblock \showarticletitle{Vocal delivery of audio description by genre: measuring users’ presence}.
\newblock \bibinfo{journal}{\emph{Perspectives}} \bibinfo{volume}{26}, \bibinfo{number}{1} (\bibinfo{year}{2018}), \bibinfo{pages}{69--83}.
\newblock


\bibitem[Wang et~al\mbox{.}(2021)]%
        {wang2021toward}
\bibfield{author}{\bibinfo{person}{Yujia Wang}, \bibinfo{person}{Wei Liang}, \bibinfo{person}{Haikun Huang}, \bibinfo{person}{Yongqi Zhang}, \bibinfo{person}{Dingzeyu Li}, {and} \bibinfo{person}{Lap-Fai Yu}.} \bibinfo{year}{2021}\natexlab{}.
\newblock \showarticletitle{Toward Automatic Audio Description Generation for Accessible Videos}. In \bibinfo{booktitle}{\emph{Proceedings of the 2021 CHI Conference on Human Factors in Computing Systems}} (Yokohama, Japan) \emph{(\bibinfo{series}{CHI '21})}. \bibinfo{publisher}{Association for Computing Machinery}, \bibinfo{address}{New York, NY, USA}, Article \bibinfo{articleno}{277}, \bibinfo{numpages}{12}~pages.
\newblock
\showISBNx{9781450380966}
\urldef\tempurl%
\url{https://doi.org/10.1145/3411764.3445347}
\showDOI{\tempurl}


\bibitem[Wu et~al\mbox{.}(2009)]%
        {wu2009biogps}
\bibfield{author}{\bibinfo{person}{Chunlei Wu}, \bibinfo{person}{Camilo Orozco}, \bibinfo{person}{Jason Boyer}, \bibinfo{person}{Marc Leglise}, \bibinfo{person}{James Goodale}, \bibinfo{person}{Serge Batalov}, \bibinfo{person}{Christopher~L Hodge}, \bibinfo{person}{James Haase}, \bibinfo{person}{Jeff Janes}, \bibinfo{person}{Jon~W Huss}, {et~al\mbox{.}}} \bibinfo{year}{2009}\natexlab{}.
\newblock \showarticletitle{BioGPS: an extensible and customizable portal for querying and organizing gene annotation resources}.
\newblock \bibinfo{journal}{\emph{Genome biology}} \bibinfo{volume}{10}, \bibinfo{number}{11} (\bibinfo{year}{2009}), \bibinfo{pages}{1--8}.
\newblock
\urldef\tempurl%
\url{https://doi.org/10.1186/gb-2009-10-11-r130}
\showDOI{\tempurl}


\bibitem[Wu et~al\mbox{.}(2022)]%
        {wu2022video}
\bibfield{author}{\bibinfo{person}{Ting Wu}, \bibinfo{person}{Junjie Peng}, \bibinfo{person}{Wenqiang Zhang}, \bibinfo{person}{Huiran Zhang}, \bibinfo{person}{Shuhua Tan}, \bibinfo{person}{Fen Yi}, \bibinfo{person}{Chuanshuai Ma}, {and} \bibinfo{person}{Yansong Huang}.} \bibinfo{year}{2022}\natexlab{}.
\newblock \showarticletitle{Video sentiment analysis with bimodal information-augmented multi-head attention}.
\newblock \bibinfo{journal}{\emph{Knowledge-Based Systems}}  \bibinfo{volume}{235} (\bibinfo{year}{2022}), \bibinfo{pages}{107676}.
\newblock
\showISSN{0950-7051}
\urldef\tempurl%
\url{https://doi.org/10.1016/j.knosys.2021.107676}
\showDOI{\tempurl}


\bibitem[Yadav et~al\mbox{.}(2015)]%
        {yadav2015multimodal}
\bibfield{author}{\bibinfo{person}{Sumit~K Yadav}, \bibinfo{person}{Mayank Bhushan}, {and} \bibinfo{person}{Swati Gupta}.} \bibinfo{year}{2015}\natexlab{}.
\newblock \showarticletitle{Multimodal sentiment analysis: Sentiment analysis using audiovisual format}. In \bibinfo{booktitle}{\emph{2015 2nd International Conference on Computing for Sustainable Global Development (INDIACom)}}. \bibinfo{pages}{1415--1419}.
\newblock


\bibitem[Yoga(2020)]%
        {tut2}
\bibfield{author}{\bibinfo{person}{UTKATA Office~Chair Yoga}.} \bibinfo{year}{2020}\natexlab{}.
\newblock \bibinfo{title}{1 Minute Office Chair Yoga - Yoga at your Desk - Flow \#1}.
\newblock \bibinfo{howpublished}{\url{https://youtu.be/vgf21Tqfwwg}}.
\newblock
\newblock
\shownote{Accessed: 2023-11-6}.


\bibitem[YouDescribe(2020)]%
        {youdescribe2020}
\bibfield{author}{\bibinfo{person}{YouDescribe}.} \bibinfo{year}{2020}\natexlab{}.
\newblock \bibinfo{title}{YouDescribe Audio Description Guideline}.
\newblock \bibinfo{howpublished}{\url{https://youdescribe.org/support/tutorial}}.
\newblock
\newblock
\shownote{Accessed: 2020-11-6}.


\bibitem[YouTube(2023)]%
        {youtube}
\bibfield{author}{\bibinfo{person}{YouTube}.} \bibinfo{year}{2023}\natexlab{}.
\newblock \bibinfo{title}{YouTube}.
\newblock \bibinfo{howpublished}{\url{http://www.youtube.com}}.
\newblock
\newblock
\shownote{Accessed: 2023-09-14}.


\bibitem[Yuksel et~al\mbox{.}(2020a)]%
        {yuksel2020human}
\bibfield{author}{\bibinfo{person}{Beste~F. Yuksel}, \bibinfo{person}{Pooyan Fazli}, \bibinfo{person}{Umang Mathur}, \bibinfo{person}{Vaishali Bisht}, \bibinfo{person}{Soo~Jung Kim}, \bibinfo{person}{Joshua~Junhee Lee}, \bibinfo{person}{Seung~Jung Jin}, \bibinfo{person}{Yue-Ting Siu}, \bibinfo{person}{Joshua~A. Miele}, {and} \bibinfo{person}{Ilmi Yoon}.} \bibinfo{year}{2020}\natexlab{a}.
\newblock \showarticletitle{Human-in-the-Loop Machine Learning to Increase Video Accessibility for Visually Impaired and Blind Users}. In \bibinfo{booktitle}{\emph{Proceedings of the 2020 ACM Designing Interactive Systems Conference}} (Eindhoven, Netherlands) \emph{(\bibinfo{series}{DIS '20})}. \bibinfo{publisher}{Association for Computing Machinery}, \bibinfo{address}{New York, NY, USA}, \bibinfo{pages}{47–60}.
\newblock
\showISBNx{9781450369749}
\urldef\tempurl%
\url{https://doi.org/10.1145/3357236.3395433}
\showDOI{\tempurl}


\bibitem[Yuksel et~al\mbox{.}(2020b)]%
        {yuksel2020increasing}
\bibfield{author}{\bibinfo{person}{Beste~F. Yuksel}, \bibinfo{person}{Soo~Jung Kim}, \bibinfo{person}{Seung~Jung Jin}, \bibinfo{person}{Joshua~Junhee Lee}, \bibinfo{person}{Pooyan Fazli}, \bibinfo{person}{Umang Mathur}, \bibinfo{person}{Vaishali Bisht}, \bibinfo{person}{Ilmi Yoon}, \bibinfo{person}{Yue-Ting Siu}, {and} \bibinfo{person}{Joshua~A. Miele}.} \bibinfo{year}{2020}\natexlab{b}.
\newblock \showarticletitle{Increasing Video Accessibility for Visually Impaired Users with Human-in-the-Loop Machine Learning}. In \bibinfo{booktitle}{\emph{Extended Abstracts of the 2020 CHI Conference on Human Factors in Computing Systems}} (Honolulu, HI, USA) \emph{(\bibinfo{series}{CHI EA '20})}. \bibinfo{publisher}{Association for Computing Machinery}, \bibinfo{address}{New York, NY, USA}, \bibinfo{pages}{1–9}.
\newblock
\showISBNx{9781450368193}
\urldef\tempurl%
\url{https://doi.org/10.1145/3334480.3382821}
\showDOI{\tempurl}


\end{thebibliography}

\end{document}